\documentclass[journal]{IEEEtran}

\makeatletter
\long\def\@makecaption#1#2{\ifx\@captype\@IEEEtablestring%
\footnotesize\begin{center}{\normalfont\footnotesize #1}\\
{\normalfont\footnotesize\scshape #2}\end{center}%
\@IEEEtablecaptionsepspace
\else
\@IEEEfigurecaptionsepspace
\setbox\@tempboxa\hbox{\normalfont\footnotesize {#1.}~~ #2}%
\ifdim \wd\@tempboxa >\hsize%
\setbox\@tempboxa\hbox{\normalfont\footnotesize {#1.}~~ }%
\parbox[t]{\hsize}{\normalfont\footnotesize \noindent\unhbox\@tempboxa#2}%
\else
\hbox to\hsize{\normalfont\footnotesize\hfil\box\@tempboxa\hfil}\fi\fi}
\makeatother

\usepackage{dblfloatfix}

\usepackage[noadjust]{cite}
\usepackage{cite}
\usepackage{amsmath,amssymb,amsfonts,dsfont}
\usepackage{algorithmic}
\usepackage{graphicx}
\usepackage{textcomp}

\usepackage{xcolor}
\usepackage[Symbolsmallscale]{upgreek}
\usepackage{enumerate}
\usepackage{comment}
\usepackage[caption=false]{subfig}
\newcommand{\qedsymbol}{\hfill\square}
\usepackage{etoolbox}
\usepackage{floatrow}
\usepackage{capt-of}

\usepackage{xr}
\newcommand{\R}{\mathbb{R}}
\newcommand{\N}{\mathbb{N}}

\newcommand{\Char}{\mathds{1}}
\newcommand{\ones}{\mathbf{1}}

\newcommand{\bfdelta}{\boldsymbol \delta}

\newtheorem{theorem}{Theorem}[section]

\newtheorem{corollary}[theorem]{Corollary}
\newtheorem{proposition}[theorem]{Proposition}
\newtheorem{lemma}[theorem]{Lemma}

\newtheorem{definition}[theorem]{Definition}
\newtheorem{remark}[theorem]{Remark}

\newcommand{\mM}{\mathbf{M}}
\newcommand{\mK}{\mathbf{K}}
\newcommand{\mL}{\mathbf{L}}

\newcommand{\mD}{\mathbf{D}}

\newcommand{\mJ}{\mathbf{J}}

\newcommand{\mP}{\mathbf{P}}

\newcommand{\mS}{\mathbf{S}}
\newcommand{\mA}{\mathbf{A}}
\newcommand{\mB}{\mathbf{B}}
\newcommand{\mC}{\mathbf{C}}

\newcommand{\mU}{\mathbf{U}}
\newcommand{\mV}{\mathbf{V}}

\newcommand{\eye}{\mathbf{I}}

\newcommand{\0}{\mathbf{0}}

\newcommand{\vx}{\mathbf{x}}
\newcommand{\vy}{\mathbf{y}}

\newcommand{\vr}{\mathbf{r}}
\newcommand{\vs}{\mathbf{s}}
\newcommand{\vecv}{\mathbf{v}}
\newcommand{\vw}{\mathbf{w}}
\newcommand{\vu}{\mathbf{u}}

\newcommand{\prob}{\mathbb{P}}

\newcommand{\cG}{\mathcal{G}}
\newcommand{\cN}{\mathcal{N}}

\newcommand{\cE}{\mathcal{E}}

\begin{document}

\title{Convergence of Bi-Virus Epidemic Models with Non-Linear Rates on Networks - A Monotone Dynamical Systems Approach

\author{Vishwaraj Doshi, Shailaja Mallick, and Do Young Eun
\thanks{{\rule{5cm}{2pt}}
\newline

Accepted for publication at IEEE/ACM Transactions on Networking, in September 2022. A subset of the material in this paper appears in \cite{vdoshi2021}.

Vishwaraj Doshi is with the Operations Research Graduate Program, Shailaja Mallick is with the Department of Computer Science, and Do Young Eun is with the Department of Electrical and Computer Engineering, North Carolina State University, Raleigh, NC. Email: \{vdoshi, smallic, dyeun\}@ncsu.edu. This work was supported in part by National Science Foundation under Grant Nos. CNS-2007423 and CNS-1824518.
}
}
}

\maketitle
\begin{abstract}
We study convergence properties of competing epidemic models of the \textit{Susceptible-Infected-Susceptible} ($SIS$) type. The SIS epidemic model has seen widespread popularity in modelling the spreading dynamics of contagions such as viruses, infectious diseases, or even rumors/opinions over contact networks (graphs). We analyze the case of \textit{two} such viruses spreading on overlaid graphs, with non-linear rates of infection spread and recovery. We call this the \textit{non-linear bi-virus model} and, building upon recent results, obtain precise conditions for \textit{global convergence} of the solutions to a \textit{trichotomy} of possible outcomes:  a virus-free state, a single-virus state, and to a coexistence state. Our techniques are based on the theory of \textit{monotone dynamical systems (MDS)}, in contrast to Lyapunov based techniques that have only seen partial success in determining convergence properties in the setting of competing epidemics. We demonstrate how the existing works have been unsuccessful in characterizing a large subset of the model parameter space for bi-virus epidemics, including all scenarios leading to coexistence of the epidemics. To the best of our knowledge, our results are the first in providing complete convergence analysis for the bi-virus system with non-linear infection and recovery rates on general graphs.
\end{abstract}  

\section{Introduction and overview}\label{introduction}

Graph-based epidemic models are widely employed to analyze the spread of real world phenomena such as communicable diseases~\cite{Yorke1976, hethcote2000mathematics}, computer viruses, malware~\cite{garetto2003modeling, yang2013epidemics, hosseini2016model}, product adoption~\cite{apt2011diffusion, prakash2012winner, ruf2017dynamics}, opinions, and rumors~\cite{trpevski2010model, zhao2013sir, lin2018opinion, Shroff2019}. The propagation of such phenomenon (which we cumulatively refer to as \textit{epidemics} or \textit{viruses}) usually takes place via processes such as human contact, word-of-mouth, exchange of emails or even in social media platforms. Graph based techniques, with edge based mechanisms to model information spread, have therefore proven to be effective in capturing such epidemic dynamics, and have been a research focus over the past few decades \cite{shakkottaiINFOCOM2014, ganesh2005effect, draief_massoulie_2009, Mieghem2013}. In recent years, the development of models which capture the competition of two or more of such epidemics has seen a surge of interest. In particular, models capturing the behavior of \textit{two competing} epidemics of the \textit{Susceptible-Infected-Susceptible} (SIS) types, also known as the \textit{bi-virus} or \textit{bi-SIS} models, have garnered significant attention over the years \cite{prakash2012winner, sahneh2014competitive, Santos2015, yang2017bi, liu2019analysis}.

Epidemic models take the form of ordinary differential equations (ODEs) and their analysis involves the identification of fixed points of the system, their uniqueness properties, and ultimately showing the convergence of the solution trajectories to those fixed points. The technique via Lyapunov functions has historically been a popular method to prove convergence to fixed points and was also used in epidemiology literature to derive the convergence properties of the SIS epidemic model. The SIS model was originally introduced in \cite{Yorke1976} to capture the spread of Gonorrhea due to contact between individuals in a population, and was further developed in \cite{n-intertwined, omic2009epidemic, van2009virus, gray2011stochastic, li2012susceptible, wang2012global, guo2013epidemic, benaim1999differential}. The central result for SIS epidemics, originally proved using Lyapunov functions in \cite{Yorke1976}, is a \textit{dichotomy} arising from the relation between model parameter ($\tau\!>\!0$) representing the effective infection rate or strength of the virus,\footnote{$\tau = \beta/\delta$, where $\beta>0$ stands for the infection rate of the virus and $\delta>0$ the recovery rate from the virus. Section \ref{epidemic models} provides a detailed explanation.} and a threshold value ($\tau^*\!>\!0$). When $\tau \!\leq\! \tau^*$, the virus spread is not strong enough and the system converges to a `virus-free' state. When $\tau \!>\! \tau^*$, it converges to a state where the virus infects a non-zero portion of the population. Attempts have also been made to perform similar convergence analysis for the bi-virus epidemic model \cite{prakash2012winner, Santos2015, yang2017bi, liu2019analysis}. The key questions posed in such literature are: Can both competing epidemics coexist over the network? If not, which one prevails? Or do both die out? This \textit{trichotomy} of possible results is what the recent literature has been trying to characterize.

When the propagation of the two epidemics occurs over the same network \cite{prakash2012winner, wang2012dynamics}, it has been established that coexistence of two viruses is impossible except in the rare cases where their effective strengths ($\tau_1,\tau_2\!>\!0$ for viruses 1, 2, respectively) are equal \cite{liu2019analysis, prakash2012winner, yang2017bi, Santos2015, sahneh2014competitive}; the virus with the larger effective strength otherwise wiping out the other, a phenomenon sometimes referred to as \textit{winner takes all} \cite{prakash2012winner}. The situation is much more complicated when the two viruses spread over two distinct networks overlaid on the same set of nodes. This modeling approach is more representative of the real world, where competing rumors/products/memes may not use the same platforms to propagate, though they target the same individuals. Recent works \cite{sahneh2014competitive, liu2019analysis, Santos2015, yang2017bi, R3, R4, R5, R6} therefore consider this more general setting, but unfortunately, a complete characterization of the trichotomy of outcomes has still proven to be elusive and remains open as of now.

While the original SIS model introduced in \cite{Yorke1976} had the aggregate infection and recovery rates of a node as \textit{linear} functions of the number of infected neighbors, there has been a push towards studying more generalized models where these rates are made heterogeneous (across nodes) and \emph{non-linear} \cite{Bansal2007, Hochberg1991, Hu2013, Barlow2000,gan_2013_computer_viruses}.
Realistic assumptions such as infection rates tending to saturation with continual increase in neighborhood infection \cite{Liu1987,yang_2015_impact,yuan_2012_modeling,ruan_2013_dynamical} have become more commonplace, implying that the models employing strictly linear spreading dynamics often provide overestimates to the real world infection rates \cite{van2009virus,yang2017bi}. This paper does not concern itself with answering which non-linear infection rate best captures the exact dynamics, but we direct the readers to \cite{yang2017bi} which provides simulation results comparing non-linear rate functions to the exact Markovian dynamics for some special randomly generated graph topologies. In some special cases, non-linear recovery rates also have an interpretation linking them to reliability theory in the form infection duration with increasing failure rates (\textit{failure} here being the recovery of an infected node). Allowing for non-linear infection and recovery rates leads to a more general version of the bi-virus model on overlaid graphs, albeit much more complicated, and the complete convergence criterion is yet to be fully established  \cite{Santos2015, yang2017bi}. It should be noted that while we extensively refer to the infection and recovery rates being either linear or non-linear in this paper, the bi-virus epidemic model itself will always be a system of non-linear ODEs.

\smallskip
\subsubsection*{\textbf{Limitations of existing works}}
Of all the recent works concerning the spread of SIS type bi-virus epidemics on overlaid networks, \cite{yang2017bi} and \cite{Santos2015} provide conditions under which the system globally converges to the state where one virus survives while the other dies out. \cite{yang2017bi} approaches the problem of showing global convergence by employing the classic technique via Lyapunov functions. However, finding appropriate Lyapunov functions is a highly non-trivial task, and as mentioned in \cite{Santos2015}, is even more difficult due to the coupled nature of the bi-virus ODE system. This can be seen in the condition they derive in \cite{yang2017bi} for the case where, say, Virus 1 dies out and Virus 2 survives. When $\tau_1$ and $\tau_2$ represent the effective strengths of Virus 1 and Virus 2, respectively, their condition translates to $\tau_1 \!\leq\! \tau_1^*$ where $\tau_1^*$ is the threshold corresponding to the single-virus case, meaning that Virus 1 would not have survived even if it was the only epidemic present on the network. More importantly, \cite{yang2017bi} is unable to characterize convergence properties for $\tau_1\!>\!\tau_1^*$ and $\tau_2\!>\!\tau_2^*$.

The authors in \cite{Santos2015} take a different approach and tackle this problem by applying their `qualitative analysis' technique, which uses results from other dynamical systems that bound the solutions of the bi-virus ODE; and provide conditions under which the system globally converges to single-virus equilibria. As we show later in Section \ref{convergence and coexistence}, however, their conditions not only characterize just a \textit{subset} of the actual space of parameters that lead to global convergence to the single-virus equilibria (which they themselves pointed out), but the size of this subset is highly sensitive to the graph topology, often much smaller than what it should be in general. In other words, a complete characterization of the \emph{entire} space of model parameters, on which the system globally converges to one of the trichotomic states, has still been recognized as an open problem in the bi-virus literature \cite{yang2017bi, Santos2015, liu2019analysis}.

\smallskip
\subsubsection*{\textbf{Our contributions}}
In this paper, we analyze the bi-virus model with \emph{non-linear} infection and recovery rates (or the \emph{non-linear bi-virus model} in short) and provide the complete characterization of the trichotomy of the outcomes with necessary and sufficient conditions under which the system globally converges to one of the three possible points: (i) a `virus-free' state, (ii) a `single-virus' equilibrium, or (iii) an equilibrium where both viruses coexist over the network. While the result for convergence to the virus-free state of the bi-SIS model is not new for non-linear infection and linear recovery rates, our proof for the same is the most general form known to date, covering the case with both infection \emph{and} recovery rates being non-linear. The proof of convergence to the virus-free state of the \textit{bi-virus} model is straightforward, and directly follows from the convergence criterion for the \textit{single-virus} SIS model with non-linear rates. However, the convergence results for fixed points where only one of the two viruses survives, or to the equilibrium where both viruses coexist, are not as straightforward to establish, rendering the typical Lyapunov based approach largely inapplicable.

In proving these results, we first show, using a specially constructed cone based partial ordering, that the bi-virus epidemic model possesses some inherent monotonicity properties. We then use novel techniques from the theory of \emph{monotone dynamical systems} (MDS) \cite{HLSmith'17} to prove our main results. In recent control systems literature \cite{DeLeenheer2001, Angeli2003, Bokharaie2010, VanHien2018, Efimov2013}, techniques based on the construction of cone based partial orderings that leverage the monotonicity properties of dynamical systems have indeed been studied. Dynamical systems exhibiting such monotonicity properties are also sometimes called deferentially positive systems \cite{Forni2016} and cooperative systems \cite{Hirsch-I} in the ODE setting, with interesting applications in consensus problems for distributed systems \cite{Altafini2013} and even neural networks \cite{Marco2012}. In this paper, we utilize these MDS techniques in the setting of competing epidemics, and as a result demonstrate an alternative to Lyapunov based approaches to analyze convergence properties of epidemic models. The novelty of using the MDS approach for analysis also lies with \cite{R2}, which uses similar techniques to analyze the bi-virus system for the special case of linear infection and recovery rates, and was developed concurrently and independently with the initial version of this work \cite{vdoshi2021}. This further highlights the utility of MDS techniques for the analysis of epidemic models on graphs.

This paper is an extension of our previous work \cite{vdoshi2021}, which gives necessary and sufficient conditions for convergence to the three types of equilibria only for the special case of the bi-virus model with \emph{linear} infection and recovery rates (or the \emph{linear bi-virus model} in short). Our conditions therein take a more precise form in terms of the model parameters $\tau_1$ and $\tau_2$ and one can visualize an exact partition of the model parameter space into regions corresponding to various convergence outcomes. We note that this partition of the model parameter space coincides with that in \cite{sahneh2014competitive}, wherein they employed only \emph{local} stability results via bifurcation analysis -- concerning only solution trajectories that originate from a small neighborhood of those fixed points. In contrast, our results in this paper concern \textit{global} stability of the system with any combination of linear as well as more general, non-linear infection and recovery rates.

\smallskip
\subsubsection*{\textbf{Structure of the paper}}
In Section \ref{epidemic models}, we first introduce the basic notation used throughout the paper, along with the classical (single-virus) SIS model and the bi-virus model. We then provide the generalization to non-linear infection and recovery rates in Section \ref{nonlinear epidemic models} with some key assumptions on the infection and recovery rate functions, complimented by a discussion in Appendix \ref{connection to DFR} regarding a special class of recovery rates. In Section \ref{monotonicity of epidemic models}, we provide a primer to the MDS theory, and establish monotonicity results for the single-virus SIS model, proving the convergence result for the single-virus model with non-linear infection and recovery rates whose proofs are deferred to Appendix \ref{single virus proofs}. We then go on to show in Section \ref{monotonicity bi-virus} that the non-linear bi-virus model is also a monotone dynamical system with respect to a specially constructed cone-based partial ordering, and include the main convergence results in Section \ref{convergence and coexistence}. In Section \ref{discussion linear} we take the opportunity to provide a more intuitive version of our results by considering the special case of linear infection and recovery rates, along with brief comparisons with the existing literature. In Section \ref{numerical results}, we provide numerical results which confirm our theoretical findings. We then conclude in Section \ref{conclusion}.

For better readability of the paper, all technical proofs of the main results are deferred to Appendix \ref{proof of the results}. The appendices also include some selected definitions and results from matrix theory (Appendix \ref{matrix theory results}), ODE theory (Appendix \ref{ODE results}), and from MDS theory (Appendix \ref{MDS}), which we use as part of our proofs of the Theorems in Section \ref{convergence and coexistence}.

\section{Preliminaries} \label{epidemic models}

\subsection{Basic Notations}
We standardize the notations of vectors and matrices by using lower case, bold-faced letters to denote vectors ($\vecv \!\in\! \R^{N}$), and upper case, bold-faced letters to denote matrices ($\mM \!\in\! \R^{N \times N}$). We denote by $\lambda(\mM)$ the largest \textit{real part}\footnote{We use the $\lambda$ notation instead of something like $\lambda_{Re}$, since it will mostly be used in cases where the largest eigenvalue is real, for which $\lambda$ itself is the largest real eigenvalue. For example, $\lambda(\mA)$ becomes the spectral radius for any non-negative matrix $\mA$ \cite{meyer_textbook, Berman1994book}.} of all eigenvalues of a square matrix $\mM$. We use $\text{diag}(\vecv)$ or $\mD_{\vecv}$ to denote the $N \!\!\times\!\! N$ diagonal matrix with entries of vector $\vecv \in \R^N$ on the diagonal. Also, we denote $\ones \!\triangleq\! [1,\!\cdots\!,1]^T$ and $\0 \!\triangleq\! [0,\!\cdots\!,0]^T$, the $N$-dimensional vector of all ones and zeros, respectively. For vectors, we write $\vx\!\leq\!\vy$ to indicate that $x_i \!\leq\! y_i$ for all $i$; $\vx \!<\! \vy$ if $\vx\!\leq\!\vy$ and $\vx\!\neq\!\vy$; $\vx \!\ll\! \vy$ when all entries satisfy $x_i \!<\! y_i$. We use $\cG(\cN,\cE)$ to represent a general, undirected, connected graph with $\cN \triangleq \{1,2,\cdots, N\}$ being the set of nodes and $\cE$ being the set of edges. When we refer to a matrix $\mA \!=\! [a_{ij}]$ as the adjacency matrix of some graph $\cG(\cN,\cE)$, it satisfies $a_{ij} \triangleq \Char_{\lbrace(i,j) \in \cE\rbrace}$ for any $i,j \in \cN$; we use $d_{min}(\mA)$ and $d_{max}(\mA)$ to denote the minimum and maximum degrees of the nodes of the corresponding graph. Since we only consider connected graphs, all the adjacency matrices in this paper are automatically considered to be irreducible (see Definition \ref{irreducible matrix} in Appendix \ref{matrix theory results}).
\subsection{$SIS$ Model with Linear rates}\label{sis epidemic model}
Consider the graph $\cG(\cN,\cE)$, and assume that at any given time $t \geq 0$, each node $i \in \cN$ of the graph is either in an \emph{infected (I)}, or in a \emph{susceptible (S)} state. An infected node can infect each of its susceptible neighbors with rate $\beta>0$.\footnote{We say an event occurs with some \emph{rate} $\alpha>0$ if it occurs after a random amount of time, exponentially distributed with parameter $\alpha>0$.} It can also, with rate $\delta>0$, be cured from its infection and revert to being susceptible again. We write $\vx(t) = [x_i(t)] \in \R^N$, where $x_i(t)$ represents the probability that node $i\in\cN$ is infected at any given time $t\geq0$. Then, the dynamics of the $SIS$ model can be captured via the system of ODEs given by
\begin{equation}\label{ode-sis-i}
    \frac{d x_i(t)}{dt} \triangleq \beta (1-x_i(t)) \sum_{j \in \cN} a_{ij} x_j(t) - \delta x_i(t)
\end{equation}
\noindent for all $i \in \cN$ and $t \geq 0$. In a matrix-vector form, this can be written as
\begin{equation}\label{ode-sis}
  \frac{d\vx}{dt} \triangleq \beta \text{diag} (\ones - \vx)\mA\vx - \delta \vx
\end{equation}
\noindent where we suppress the $(t)$ notation for brevity. The system \eqref{ode-sis} is positively invariant in the set $[0,1]^N$, and has $\0$ as a fixed point (the virus-free equilibrium). The following result is well known from \cite{Yorke1976}, which we will generalize in Section \ref{monotonicity and convergence sis}.

\begin{theorem}[Theorem 3.1 in \cite{Yorke1976}] \label{sis-conditions}
  Let $\tau \!\triangleq\! \beta/\delta$. Then,
  \begin{enumerate}[(i)]
    \item either $\tau\leq 1/\lambda(\mA)$, and $\vx^* = \0$ is a globally asymptotically stable fixed point of \eqref{ode-sis};
    \item or $\tau > 1/\lambda(\mA)$, and there exists a unique, strictly positive fixed point $\vx^* \in (0,1)^N$ such that $\vx^*$ is globally asymptotically stable in $[0,1]^N\setminus \{\0\}$.$\qedsymbol$
  \end{enumerate}
\end{theorem}

\subsection{Bi-Virus Model with Linear rates}\label{bi-virus epidemic model}

Consider two graphs $\cG_1(\cN,\cE_1)$ and $\cG_2(\cN,\cE_2)$, on the same set of nodes $\cN$ but with different edge sets $\cE_1$ and $\cE_2$. At any given time $t\geq 0$, a node $i \in \cN$ is either \textit{infected by Virus 1}, \textit{infected by Virus 2}, or is \textit{susceptible}. A node infected by Virus 1 infects each of its susceptible neighbors with rate $\beta_1 > 0$, just like in the $SIS$ model, but does so only to nodes which are its neighbors with respect to the graph $\cG_1(\cN,\cE_1)$. Nodes infected by Virus 1 also recover with rate $\delta_1>0$, after which they enter the susceptible state. Similarly, nodes infected by Virus 2 infect their susceptible neighbors, this time with respect to the graph $\cG_2(\cN,\cE_2)$, with rate $\beta_2>0$, while recovering with rate $\delta_2>0$. This competing bi-virus model of epidemic spread, also referred to as the $SI_1I_2S$ model, can be represented by the following ODE system:
\begin{equation}\label{eq:biSIS-i}
\begin{split}
    \frac{d x_{i}}{dt} &\triangleq \beta_1 \left(1- x_i - y_i\right) \sum_{j \in \cN} a_{ij} x_{j} - \delta_1 x_{i} \\
    \frac{d y_{i}}{dt} &\triangleq \beta_2 \left(1- x_i - y_i\right) \sum_{j \in \cN} b_{ij} y_{j} - \delta_2 y_{i}
\end{split}
\end{equation}
\noindent for all $i \in \cN$ and $t\geq 0$. In matrix-vector form, \eqref{eq:biSIS-i} becomes:
\begin{equation}\label{eq:biSIS}
\begin{split}
    \frac{d \vx}{dt} &\triangleq \beta_1 \text{diag}\left(\ones - \vx - \vy\right) \mA \vx - \delta_1 \vx \\
    \frac{d \vy}{dt} &\triangleq \beta_2 \text{diag}\left(\ones - \vx - \vy\right) \mB \vy - \delta_2 \vy,
\end{split}
\end{equation}
\noindent where $\mA = [a_{ij}]$ and $\mB = [b_{ij}]$ are the adjacency matrices of graphs $\cG_1(\cN,\cE_1)$ and $\cG_2(\cN,\cE_2)$, respectively.

\section{Epidemic Models with Non-linear Infection and Recovery rates} \label{nonlinear epidemic models}

In this section, we introduce the single-virus and bi-virus SIS models with non-linear infection and recovery rates. Non-linearities can be attributed to the spread and recovery from the virus being related to the susceptibility of the disease (or its prevalence in the population) in a more complicated manner. This is more general than simply exponential random variables with constant rates used to model the spreading and recovery processes, which in aggregate scale linearly with the infection probabilities.\footnote{`Aggregate' here refers to the mean field approximation which is one way to derive SIS-type ODEs. Another way is the large population mean field \textit{limit} of a stochastic process, where the connection to the corresponding ODE system is formed via the Kurtz's theorem \cite{draief_massoulie_2009}. In this case, linearity is induced by the uniform or homogeneous mixing assumption which is also a subject of criticism in epidemiology literature \cite{Bansal2007, Hochberg1991, Hu2013, Barlow2000}.} This  is shown to be limiting in accurately modelling the trajectories of an infection spread; the linear scaling of the infection and recovery rates shown to being an overestimate to what is observed in reality \cite{yang2017bi, Hu2013}. Many works thus argue for the modelling of these spreading processes with non-linear functions \cite{Barlow2000, Bansal2007, Hochberg1991, Liu1987}. We first present the more general single-virus SIS model with a set of intuitive assumptions (A1)--(A5) for the non-linear infection and recovery rates.

\subsection{$SIS$ Model with Non-linear rates}

In \eqref{ode-sis-i} the term $\sum_{j \in \cN} a_{ij}x_j(t)$ denotes the overall rate at which a susceptible node $i \in \cN$ gets infected by its neighbors. In what follows, we replace this by a generic function $f_i(\vx(t))$, thereby allowing the overall infection rate for each node to be any non-linear function of $x_j(t)$ for all neighbors $j$ of $i$. Similarly, we replace the term $\delta x_i(t)$, denoting the overall recovery rate for any node $i\in \cN$, by a non-linear function $q_i(\vx(t))$. This generic version of the SIS model, allowing for non-linear infection and recovery rates, is given by the ODE
\begin{equation}\label{ode-sis-i-nonlinear}
    \frac{d x_i(t)}{dt} = \bar f_i(\vx(t))  \triangleq (1-x_i(t)) f_i(\vx(t)) - q_i(\vx(t))
\end{equation}
\noindent for all $i \in \cN$ and $t \geq 0$. In a matrix-vector form, this can be written as
\begin{equation}\label{ode-sis-nonlinear}
  \frac{d\vx}{dt} = \bar F(\vx) \triangleq \text{diag} (\ones - \vx)F(\vx) - Q(\vx)
\end{equation}
\noindent where $F(\vx) = [f_i(\vx)]\in \R^N$, and $Q(\vx) = [q_i(\vx)] \in \R^N$ are the vectors of non-linear infection and recovery rate functions, respectively. We assume that they are continuous and twice differentiable in $[0,1]^N$, with $\mJ_F(\vx)$ and $\mJ_Q(\vx)$ denoting the Jacobians of $F$ and $Q$ respectively, evaluated at any point $\vx \in [0,1]^N$. We now make the following key assumptions:

\smallskip
\begin{itemize}
  \item[(A1)] $F(\0) = \0$ and $Q(\0) = \0$;
  \smallskip
  \item[(A2)] $\left[\mJ_F(\vx)\right]_{ij} = \frac{\partial f_i (\vx)}{\partial x_j} > 0 ~~\forall i\neq j$ with $a_{ij}>0$, otherwise $\left[\mJ_F(\vx)\right]_{ij} = 0$;
  \smallskip
  \item[(A3)] $\left[\mJ_Q(\vx)\right]_{ii} \!=\! \frac{\partial q_i (\vx)}{\partial x_i} \!>\! 0$, and $\left[\mJ_Q(\vx)\right]_{ij} \!=\! \frac{\partial q_i (\vx)}{\partial x_j} \!\leq\! 0$ for all $i \!\neq\! j$, $\vx \in [0,1]^N$. Moreover, $\sum \limits_{j \neq i} \left[\mJ_Q(\vx)\right]_{ij} \!<\! \left[\mJ_Q(\vx)\right]_{ii}$;
  \smallskip
  \item[(A4)] $f_i(\vx)$ is concave in $[0,1]^N\!\!$, that is, $\frac{\partial^2f_i}{\partial x_j\partial x_k} \!\leq\! 0$ for all $i,\!j,\!k \!\in\! \cN$;
  \smallskip
  \item[(A5)] $q_i(\vx)$ is convex function of $x_i \in [0,1]^N$, and a concave function of $x_j$ for all $j \neq i$. That is, $\frac{\partial^2q_i}{\partial^2x_i} \geq 0$ and $\frac{\partial^2 q_i}{\partial x_j \partial x_k} \leq 0$ for all $i \!\in\! \cN$, and $j,k \!\in\! \cN \!\setminus\! \{ i \}$.
\end{itemize}
\smallskip

Assumption (A1) ensures that the virus-free state is a fixed point of \eqref{ode-sis-nonlinear}, while (A2) is a proximity assumption that models infection spread only through edges of the underlying graph. Assumption (A3) concerns with the recovery rate, allowing it to be reduced by infected neighbors while still being no-negative. (A4) and (A5) assume concavity properties of the functions $f_i(\vx)$ and $q_i(\vx)$ in $x_j$ for any neighbor $j$ of $i$. This allows the effect of neighborhood infection $x_j$ to \textit{saturate}\footnote{As $x_j$ increases for any neighbor $j$ of node $i$, the magnitude of the \textit{resulting change} in both infection rate $f_i(\vx)$ and recovery rate $q_i(\vx)$ decreases. This is similar to the case of diminishing returns.} as $x_j$ increases. Assumption (A5) also assumes convexity of $q_i(\vx)$ in \textit{local} infection $x_i$, which means that increase in recovery rate caused by $x_i$ can be larger as $x_i$ increases.

Examples for non-linear infection rates satisfying (A1)--(A5) include logarithmic functions $f_i(\vx) = \sum_{j}a_{ij}\ln{(1+x_j)}$, similar to those in \cite{yang2017bi}. Examples of non-linear recovery rates include polynomial functions such as $q_i(\vx) = (1+x_i)^k - 1$ for any $k\geq1$. A special class of the permissible non-linear recovery rates, where the infection duration is dependent solely on local infection $x_i$, is related to processes that have \textit{decreasing failure rates}  (DFR)\footnote{\textit{Failure rate} for a non-negative random variable is defined as the ratio between its probability density function (PDF) and its complimentary cumulative distribution function (CCDF). In the context of infection duration, \textit{decreasing failure rate} means that nodes recover at a decreased rate the longer they stay continuously infected. A more detailed discussion regarding the connection to SIS recovery rates can be found in Appendix \ref{connection to DFR}.}. This special class of recovery processes that are DFR also includes the case of linear recovery rates. Note that our assumptions allow $f_i(\vx)$ and  $q_i(\vx)$ to be \textit{heterogeneous} across all nodes $i \in \cN$, and the case with linear rates in \eqref{ode-sis} readily satisfies (A1)--(A5). This also extends to the linear bi-virus model \eqref{eq:biSIS} being a special case of the non-linear bi-virus model introduced in the next subsection, with infection and recovery rate functions therein satisfying the same assumptions (A1)--(A5).

\begin{figure}[!t]
    \centering
    {\includegraphics[scale=0.65]{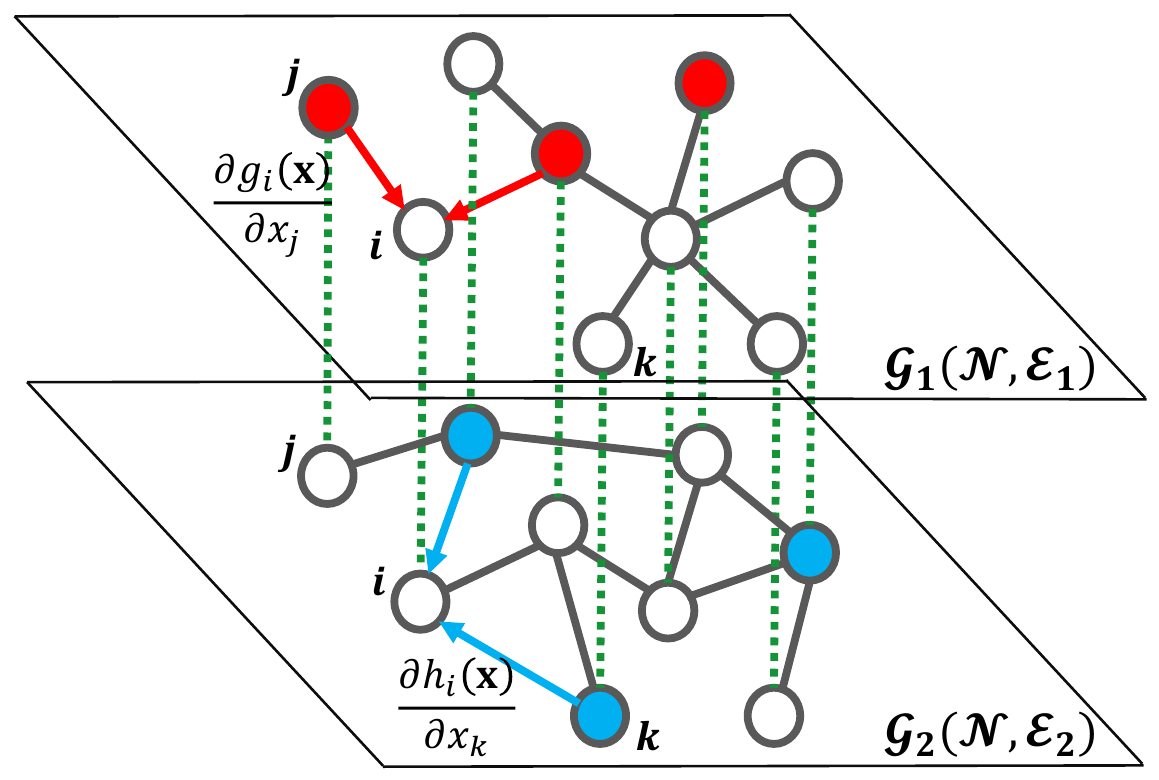}}
    \caption{Bi-Virus epidemic spread across overlaid graphs sharing the same set of nodes. Red and Blue arrows denote the spread of Virus 1 and 2, respectively from infected nodes $j$ and $k$ (coloured Red and Blue) to the susceptible node $i$ (uncoloured) with the instantaneous rates as shown. The infected Red and Blue nodes also recover with a total rate of $r_i(\vx)$ and $s_i(\vy)$ for any node $i\in \cN$, respectively.}\vspace{-2mm}
    \label{fig:bi-virus spread}
\end{figure}

\subsection{Bi-Virus Model with Non-linear rates}

The Bi-Virus model with non-linear infection and recovery rates is given by the following coupled system of ODEs:
\begin{equation}\label{eq:biSIS-i-nonlinear}
\begin{split}
    \frac{d x_{i}}{dt} = \bar g_i(\vx,\vy)&\triangleq \left(1- x_i - y_i\right) g_i(\vx(t)) - r_i(\vx) \\
    \frac{d y_{i}}{dt} = \bar h_i(\vx,\vy)&\triangleq \left(1- x_i - y_i\right) h_i(\vy(t)) - s_i(\vy)
\end{split}
\end{equation}
\noindent for all $i \in \cN$ and $t\geq 0$. In a matrix-vector form, \eqref{eq:biSIS-i-nonlinear} becomes:
\begin{equation}\label{eq:biSIS-nonlinear}
\begin{split}
    \frac{d \vx}{dt} = \bar G(\vx,\vy) &\triangleq \text{diag}\left(\ones - \vx - \vy\right) G(\vx) - R(\vx) \\
    \frac{d \vy}{dt} = \bar H(\vx,\vy) &\triangleq \text{diag}\left(\ones - \vx - \vy\right) H(\vy) - S(\vy),
\end{split}
\end{equation}
Where $G(\vx) = [g_i(\vx)]$, $R(\vx) = [r_i(\vx)]$, and $H(\vy) = [h_i(\vy)]$, $S(\vy) = [s_i(\vy)]$ are the non-linear infection and recovery rate functions for viruses 1 and 2, respectively. The pairs $(G,R)$ and $(H,S)$ each satisfy the assumptions (A1)--(A5); where $G$ and $H$ specifically satisfy (A2) with respect to their corresponding graphs with adjacency matrices $\mA$ and $\mB$, respectively. Figure \ref{fig:bi-virus spread} illustrates of how these competing epidemics spread over the corresponding overlaid graphs.

Assumptions (A1)--(A5) are also more general (weaker) than those assumed in \cite{Santos2015, yang2017bi}, where the recovery rates are restricted to being linear functions and are thus a special case of our model. We emphasize that while the set off assumptions for non-linear rates are mostly similar to (slightly more general than) those in literature, the characterization of all convergence scenarios for their respective bi-virus models is incomplete, as we shall discuss later in Section \ref{discussion linear}.

\section{Monotone Dynamical Systems and the Single Virus Epidemic}\label{monotonicity of epidemic models}

In this section, we provide a succinct introduction to monotone dynamical systems (MDS) and some important definitions therein. We go on to show that the $SIS$ model \eqref{ode-sis-nonlinear} is a monotone dynamical system (specifically a cooperative system) and briefly apply these MDS techniques to epidemic models by deriving the exact convergence result of the non-linear $SIS$ model. We also observe that Theorem \ref{sis-conditions} is a special case for when the infection and recovery rates are linear.

\subsection{Monotone Dynamical Systems - A Primer}\label{primer}
A well known result from real analysis is that monotone sequences in compact (closed and bounded) subsets of $\R^n$ converge in $\R^n$ \cite{Analysis-Yeh}. This simple, yet powerful result has been fully integrated with the theory of dynamical systems in a series of works \cite{Hirsch-I,Hirsch-II,Hirsch-III,Hirsch-IV,Hirsch-V,HLSmith'88,HLSmith'90,HLSmith'91,HLSmith'04, MDSbook}, which cumulatively form the theory of \textit{monotone dynamical systems} (MDS). The foundations of MDS were laid down in \cite{Hirsch-I,Hirsch-II,Hirsch-III,Hirsch-IV,Hirsch-V} which study ordinary differential equations, specifically \emph{cooperative} ODE systems. We here provide a brief, informal introduction to such ODE systems, with more details in Appendix \ref{MDS}.

A central tool in the theory of MDS is the notion of \textit{generalized cone-orderings}, which extends the concept of monotonicity in vector spaces.
\smallskip
\begin{definition}
  Given a convex cone $K \subset X$ for any vector space $X$, the \textit{cone-ordering} $\leq_K$ ($<_K$, $\ll_K$) generated by $K$ is an order relation that satisfies
  \begin{enumerate}[(i)]
    \item
    $~\vx \!\leq_K\! \vy \!\iff\! (\vy\!-\!\vx) \in K$;
    \item
    $~\vx \!<_K\! \vy \!\iff\! \vx \!\leq_K\! \vy$ and $\vx \!\neq\! \vy$; and
    \item
    $~\vx \!\ll_K\! \vy \!\iff\! (\vy\!-\!\vx) \in \text{int}(K)$, for any $\vx,\vy \in X$.
  \end{enumerate}
\end{definition}\smallskip
Note that, `$\ll_K$' implies `$<_K$' and is a stronger relation. Cone-orderings generated by the positive orthant $K \!=\! \R^n_+$ are simply denoted by $\leq$ ($<, \ll$), that is, without the `$K$' notation.

Let $\phi_t(\vx)$ denote the solution of a dynamical system at some time $t \!>\! 0$ starting from an initial point $\phi_0(\vx) \!=\! \vx \!\in\! \R^n$.
\smallskip
\begin{definition}
  Given a cone-ordering $\leq_K$ ($<_K$, $\ll_K$), the dynamical system is said to be \emph{monotone} if for every $\vx,\vy \!\in\! \R^n$ such that $\vx \!\leq_K\! \vy$, we have $\phi_t(\vx) \!\leq_K\! \phi_t(\vy)$ for all $t \!>\! 0$. The system is called \emph{strongly monotone} if for all $\vx,\vy \!\in\! \R^n$ such that $\vx \!<_K\! \vy$, we have $\phi_t(\vx) \!\ll_K\! \phi_t(\vy)$ for all $t \!>\! 0$.
\end{definition}
\smallskip

The main result from MDS theory says that (almost) every solution trajectory of a \textit{strongly monotone} system always converges to some equilibrium point of the system \cite{Hirsch-II,HLSmith'91,HLSmith'04,HLSmith'17}. If the system has only one stable fixed point, then this in itself is enough to prove global convergence. Monotonicity properties of a dynamical system can therefore be leveraged as an alternative to constructing Lyapunov functions, which is often intractable.

Consider the following autonomous ODE system
\begin{equation}\label{ode sys}
  \dot \vx  = \bar F(\vx),
\end{equation}
\noindent where $\bar F(\vx) = [\bar f_i(\vx)] \in \R^n$ is the vector field. If $\phi_t(\vx)$ is the solution of this ODE system, we say the system is \emph{co-operative} if it is monotone. There are ways to find out whether an ODE system is co-operative or not. In particular, one can answer this by observing the Jacobian of the vector field \cite{HS_Co-Op}. The so-called \textit{Kamke condition} \cite{MDSbook} says that \eqref{ode sys} is co-operative with respect to the cone-ordering generated by the positive orthant $K=\R^n_+$ if and only if 
\begin{equation}\label{Kamke for positive}
  \frac{\partial \bar f_i}{\partial x_i} \geq 0, ~~~~~~~ \text{for all } i \neq j.
\end{equation}
\noindent While it is not straightforward to obtain such a clean condition for any general convex cone $K$, one can still deduce the co-operative property of the ODE with respect to any one of the other orthants of $\R^n$ by observing the signed entries of the Jacobian. We will show how this is done for the bi-virus system \eqref{eq:biSIS} later in Section \ref{monotonicity bi-virus}.

If the Jacobian of an ODE system is an irreducible matrix in a subset $D$ of the state space, we say that the ODE system is \textit{irreducible in $D$} (Definition \ref{irreducible ode} in Appendix \ref{MDS}). If the ODE system is co-operative in $D$ as well as irreducible in $D$, then it is strongly monotone in $D$ (Theorem \ref{SM ODE} in Appendix \ref{MDS}). To prove convergence properties, we should ideally be able to show that our system is strongly monotone in the entirety of the state space it is contained in, for which we can directly apply the main MDS convergence result. However, this is often not the case, and one needs additional results from MDS literature to prove convergence. These details are deferred to Appendix \ref{MDS}.

\subsection{Monotonicity and convergence of SIS epidemic models}\label{monotonicity and convergence sis}

The following proposition establishes the monotonicity of the single-virus SIS model with non-linear infection and recovery rates with respect to the regular ordering relationship (cone-ordering generated by $R^N_+$).
\smallskip
\begin{proposition}\label{prop: single sis cooperative}
  The ODE system \eqref{ode-sis-nonlinear} is cooperative in $[0,1]^N$ and irreducible in $(0,1)^N$ with respect to the cone-ordering generated by the positive orthant $\R^N_+$.$\qedsymbol$
\end{proposition}\smallskip

We now state the convergence criterion for the non-linear single-virus $SIS$ model.

\smallskip
\begin{theorem}\label{nonlinear sis-conditions}
  Let $\mJ_F(\vx)$ and $\mJ_Q(\vx)$ denote the Jacobian matrices of the vector valued infection and recovery rate functions $F(\vx)$ and $Q(\vx)$ from \eqref{ode-sis-nonlinear}, respectively. Then,
  \begin{enumerate}[(i)]
    \item either $\lambda(\mJ_F(\0) - \mJ_Q(\0)) \leq 0$, and $\vx^*=0$ is the globally asymptotically stable fixed point of \eqref{ode-sis-nonlinear};
    \item or $\lambda(\mJ_F(\0) - \mJ_Q(\0)) > 0$, and there exists a unique, strictly positive fixed point $\vx^*\gg0$ such that $\vx^*$ is globally asymptotically stable in $[0,1]^N\setminus \{ \0 \}$.$\qedsymbol$
  \end{enumerate}
\end{theorem}\smallskip

The proof for Theorem \ref{nonlinear sis-conditions} utilizes a result from the monotone dynamical systems literature, provided as Theorem \ref{limit set trichotomy} in Appendix \ref{single virus proofs}. It was originally proved and applied to linear SIS epidemics in \cite{Krause_trichotomy} as an alternate proof of the convergence properties of the model for Gonorrhea spread in \cite{Yorke1976}, which is a special case of our non-linear model \eqref{ode-sis-nonlinear}. We can also see this in the following remark.

\smallskip
\begin{remark}
For the single-virus SIS model with linear infection and recovery rates \eqref{ode-sis}, the conditions derived in Theorem \ref{nonlinear sis-conditions} reduce to those in Theorem \ref{sis-conditions}.
\end{remark}
\smallskip
\begin{IEEEproof}
By substituting $F(\vx) = \beta \mA \vx$ and $Q(\vx) = \delta \vx$ in \eqref{jacobian-single} (Jacobian of the single-virus system \eqref{ode-sis-nonlinear}, mentioned in the proof of Theorem \ref{nonlinear sis-conditions}) and evaluating at $\vx=\0$, we get $\mJ_{\bar F}(\0) = \mJ_{F}(\0) \!-\! \mJ_Q(\0) = \beta \mA \!-\! \delta \eye$. The condition $\lambda(\mJ_F(\0) \!-\! \mJ_Q(\0)) = \lambda(\beta \mA \!-\! \delta \eye) > 0$ $(\leq 0)$ can be rewritten as $\tau > 1/\lambda(\mA)$ $\left(\leq 1/\lambda(\mA)\right)$ where $\tau = \beta/\delta$, which as the same as in Theorem \ref{sis-conditions}.
\end{IEEEproof}
\smallskip

While Theorem \ref{nonlinear sis-conditions} could be proved using the steps in \cite{Yorke1976}, which were recreated again in \cite{yang2017bi}, it requires first the application of two different Lyapunov functions and also requires proving the uniqueness of the positive fixed point. Alternatively, one could apply  Theorem 1 in \cite{R1} to establish the uniqueness of the positive fixed point by first showing that the Jacobian of $\bar F(\vx)$ evaluated at any point $\vx \gg\0$ satisfying $\bar F(\vx) = \0$, is Hurwitz. This, combined with Proposition \ref{prop: single sis cooperative}, could then provide the necessary convergence criterion. However, we maintain that using Theorem \ref{limit set trichotomy} would be a simpler way to derive the same results, whose proof is deferred to Appendix \ref{single virus proofs}.

\section{Main results for the non-linear Bi-Virus model}

We provide the necessary and sufficient results on the non-linear infection and recovery rates of the bi-virus system \eqref{eq:biSIS-nonlinear} for convergence to each of the three different kinds of equilibria: the virus-free, the single-virus equilibrium, and the co-existence equilibrium. However, before stating the main convergence results (proofs deferred to Appendix F in \cite{Appendices}), we establish the monotonicity of the non-linear bi-virus model.

\subsection{Monotonicity of the Bi-Virus epidemic models} \label{monotonicity bi-virus}

We first revisit the Kamke condition from Section \ref{primer}, in this instance given for a the \emph{southeast cone-ordering} as stated below.

\smallskip
\subsubsection*{\textbf{Southeast cone-ordering and the Kamke condition}}
Consider the cone-ordering generated by the convex cone $K=\{ \R^N_+ \times \R^N_- \} \subset \R^{2N}$. This cone is one of the orthants of $\R^{2N}$, and for $N=1$, it would correspond to the \textit{southeast} orthant of $\R^2$ $\left( K = \{ \R_+ \times \R_- \} \subset \R^2 \right)$. For any two points $(\vx,\vy)$, $(\bar \vx, \bar \vy) \in \R^{2N}$, it satisfies the following:
\begin{enumerate}[(i)]
\item
$(\vx,\vy) \!\leq_K\! (\bar \vx,\bar \vy) \!\iff\! x_i \!\leq\! \bar x_i$ and $y_i \!\geq\! \bar y_i$ for all $i \!\in\! \cN$;
\item
$(\vx,\vy) \!\!<_K\!\! (\bar \vx,\bar \vy) \!\!\iff\!\! (\vx,\vy) \!\!\leq_K\!\! (\bar \vx,\bar \vy)$ \!and\! $(\vx,\vy) \!\neq\! (\bar \vx,\bar \vy)$;
\item
$(\vx,\vy) \!\ll_K\! (\bar \vx,\bar \vy) \!\iff\! x_i \!<\! \bar x_i$ and $y_i \!>\! \bar y_i$ for all $i \!\in\! \cN$.
\end{enumerate}
This type of cone-ordering is often referred to as the \textit{southeast cone-ordering}, and the corresponding cone $K$ is the \textit{southeast orthant} of $\R^{2N}$. As shown in \cite{HS_Co-Op}, the Kamke condition for determining whether an ODE system is cooperative or not with respect to the positive orthant $\R^{2N}_+$ can be generalised for cone-orderings generated by any orthant of $\R^{2N}$, including the southeast orthant. Once again, this is done by observing the Jacobian of the respective ODE system. Consider the $2N$ dimensional system given by
\begin{equation*}
  \dot \vx = \bar G(\vx,\vy) ~~\text{and}
  ~~\dot \vy = \bar H(\vx,\vy),
\end{equation*}
\noindent where $\bar G(\vx,\vy) = [\bar g_i(\vx,\vy)]$ and $\bar H(\vx,\vy) = [\bar h_i(\vx,\vy)]$ are vector-valued functions in $\R^N$. The Kamke condition for this system with respect to the southeast cone-ordering \cite{HS_Co-Op} is
\begin{equation*}
    \frac{\partial \bar g_i}{\partial x_j} \geq 0, ~ \frac{\partial \bar h_i}{\partial y_j} \geq 0, ~\forall i\neq j, ~~~\text{and}~~~
\frac{\partial \bar g_i}{\partial y_j} \leq 0, ~ \frac{\partial \bar h_i}{\partial x_j} \leq 0, ~\forall i,j.
\end{equation*}
\noindent Roughly speaking, the Jacobian $\mJ_{GH} (\vx,\vy)$ of the system, evaluated at all points in the state space, should be in the following block matrix form (where the signs are not strict):
\begin{equation}\label{Jacobian form}
\mJ_{\bar G \bar H}=
\begin{bmatrix}
*  &+  &+  &- &- &- \\
+  &*  &+  &- &- &- \\
+  &+  &*  &- &- &- \\
-  &-  &-  &* &+ &+ \\
-  &-  &-  &+ &* &+ \\
-  &-  &-  &+ &+ &*
\end{bmatrix}
\end{equation}
\noindent  Note that the state space of the ODE system \eqref{eq:biSIS} is given by $D \triangleq \left\{ (\vx,\vy) \in [0,1]^{2N} ~|~ \vx + \vy \leq \ones \right\}$.
\smallskip
\begin{proposition}\label{cooperative interor prop}
  The ODE system \eqref{eq:biSIS-nonlinear} (the non-linear bi-virus model) is cooperative in $D$ with respect to the southeast cone-ordering. It is also irreducible in $\text{Int}(D)$.
\end{proposition}
\smallskip
\begin{IEEEproof}
  For all $(\vx,\vy) \in D$ and $i \neq j \in \cN$, we have
  $$\frac{\partial \bar g_i(\vx,\vy)}{\partial x_j} = (1-x_i-y_i)\frac{\partial g_i(\vx)}{\partial x_j} -\frac{\partial r_i(\vx)}{\partial x_j} \geq 0,$$
  $$\frac{\partial \bar h_i(\vx,\vy)}{\partial y_j} = (1-x_i-y_i)\frac{\partial h_i(\vy)}{\partial y_j} -\frac{\partial s_i(\vx)}{\partial y_j} \geq 0$$
  \noindent since $\frac{\partial g_i(\vx)}{\partial x_j}\geq 0$, $\frac{\partial r_i(\vx)}{\partial x_j}\leq 0$  and $\frac{\partial h_i(\vy)}{\partial y_j}\geq 0$, $\frac{\partial s_i(\vy)}{\partial y_j}\leq 0$ from assumptions (A2) and (A3), and $(1-x_i-y_i)\geq 0$. Moreover for all $i \in \cN$,
  $$\frac{\partial \bar g_i}{\partial y_i} = -g_i(\vx) \leq 0
  ~~\text{and}~~
  \frac{\partial \bar h_i}{\partial x_i} = -h_i(\vy) \leq 0,$$
  \noindent with ${\partial \bar g_i}/{\partial y_j}={\partial \bar h_i}/{\partial x_j}=0$. Thus, the Kamke conditions are satisfied and the system is cooperative in $D$.

  The Jacobian $\mJ_{\bar G \bar H} (\vx,\vy)$ of system \eqref{eq:biSIS} is written as
  \smallskip
  \begin{equation}\label{Jacobian bi-virus}
    \begin{split}
      &\!\!\!\mJ_{\bar G \bar H}(\vx,\vy) = \\
      &\!\!\!\!\!\begin{bmatrix}
         \mS_{\vx\vy}\mJ_G(\vx) \!\!-\!\! \mD_{G(\vx)} \!\!-\!\! \mJ_R(\vx)      &\!\!\!\!-\! \mD_{G(\vx)}  \\
        -\! \mD_{H(\vy)}    &\!\!\!\!\mS_{\vx\vy}\mJ_H(\vy) \!\!-\!\!  \mD_{H(\vy)} \!\!-\!\! \mJ_S(\vy)
      \end{bmatrix}\!\!,
    \end{split}
  \end{equation}
  \smallskip
  \noindent where $\mS_{\vx,\vy} \triangleq \text{diag}(\ones - \vx - \vy)$, $\mD_{G(\vx)} \triangleq \text{diag}(G(\vx))$ and $\mD_{H(\vy)} \triangleq \text{diag}(H(\vy))$. Since the infection rate functions satisfy assumption (A2) for their corresponding underlying graphs, $\mJ_G(\vx)$ and $\mJ_H(\vy)$ follow the sign structure of $\mA$ and $\mB$ respectively and are irreducible. The off-diagonal blocks of $\mJ_{\bar G \bar H}(\vx,\vy)$ are diagonal matrices with non-zero diagonal entries for $(\vx,\vy) \in \text{Int}(D)$, and there does not exist a permutation matrix that would transform this into a block upper triangular matrix. Hence, by Definition \ref{irreducible ode}, the system is irreducible in $\text{Int}(D)$, and this completes the proof.
\end{IEEEproof}\smallskip

From Proposition \ref{cooperative interor prop}, we deduce that the non-linear bi-virus system of ODEs \eqref{eq:biSIS-nonlinear} is co-operative in $D$, and thus strongly monotone in $\text{Int}(D)$ in view of Theorem \ref{SM ODE} in Appendix \ref{MDS}. This property also extends to the linear bi-virus system \eqref{eq:biSIS} which is a special case of \eqref{eq:biSIS-nonlinear}.

\subsection{Convergence and Coexistence properties of the Bi-Virus model}  \label{convergence and coexistence}

We are now ready to establish results on convergence properties of the bi-virus model and provide conditions for coexistence of two viruses in the non-linear bi-virus model as in \eqref{eq:biSIS-nonlinear}.

Let $\vx^*$ and $\vy^*$ be the globally attractive fixed points of the single-virus SIS models that system \eqref{eq:biSIS-nonlinear} would reduce to when Virus 2 and 1, respectively, are not present over the network. These systems are given by
\begin{equation}\label{sis-x}
  \dot \vx = F^x(\vx) \triangleq \bar G(\vx,\0) = \text{diag}(\ones - \vx)G(\vx) - R(\vx),
\end{equation}
\begin{equation}\label{sis-y}
  \dot \vy = F^y(\vy) \triangleq \bar H(\0,\vy) = \text{diag}(\ones - \vy)H(\vy) - S(\vy);
\end{equation}
\noindent and by Theorem \ref{nonlinear sis-conditions}, $\vx^*\!=\!\0$ ($\vy^*\!=\!\0$) if $\lambda \left( \mJ_G(\0) \!-\! \mJ_R(\0) \right) \!\leq\! 0$ (if $\lambda \left( \mJ_H(\0) \!-\! \mJ_S(\0) \right) \!\leq\! 0$), and $\vx^*\!\gg\!\0$ ($\vy^*\!\gg\!\0$) otherwise.

We first state the result when the virus-free equilibrium is globally attractive. We prove this by presenting simple arguments which require only Theorem \ref{nonlinear sis-conditions} for SIS model along with the monotonicity properties derived in the previous section, eliminating the need of a Lyapunov based approach.
\smallskip
\begin{theorem}[Convergence to virus-free equilibria]\label{theorem virus free}
  If $\lambda \left( \mJ_G(\0) \!-\! \mJ_R(\0) \right) \!\leq\! 0$ and $\lambda \left( \mJ_H(\0) \!-\! \mJ_S(\0) \right) \!\leq\! 0$, trajectories of \eqref{eq:biSIS-nonlinear} starting from any point in $D$ converge to $(\0,\0)$.$\qedsymbol$
\end{theorem}\smallskip

We next characterize the conditions when the system globally converges to equilibria when only one of the viruses survives over the network. Let $\mS_\vx \!\triangleq\! \text{diag}(\ones\!-\!\vx)$ and $\mS_\vy \!\triangleq\! \text{diag}(\ones\!-\!\vy)$ for any $\vx,\vy \in \R^N$. Also denote by $B_x \!\triangleq\! \left\{ (\vx,\vy) \in D ~|~ \vx\!>\!\0 \right\}$ the set of all points $(\vx,\vy) \!\in\! D$ for which $x_i\!>\!0$ for some $i\in\N$, and let $B_y \!\triangleq\! \left\{ (\vx,\vy) \in D ~|~ \vy\!>\!\0 \right\}$ be a similar set for the $y_i$ entries.\linebreak
\smallskip
\begin{theorem}[Convergence to single-virus equilibria]\label{theorem wta}
  When $\lambda \!\left( \mS_{\vy^*}\mJ_G(\0) \!-\! \mJ_R(\0) \right) \!>\! 0$ and $\lambda \!\left( \mS_{\vx^*}\mJ_H(\0) \!-\! \mJ_S(\0) \right) \!\leq\! 0$, $(\vx^*,\0)$ is globally attractive in $B_x$;\footnote{We consider $B_x$ as the global domain of attraction instead of $D$ because $\vx=0$ for all points in the set $D\setminus B_x$. Starting from such points the system is no longer a bi-virus epidemic, but a single-virus SIS system for Virus 2.} that is, every trajectory of system \eqref{eq:biSIS-nonlinear} starting from points in $B_x$ converges to $(\vx^*,\0)$.

  Similarly, when $\lambda \left( \mS_{\vy^*}\mJ_G(\0) - \mJ_R(\0) \right) \leq 0$ and $\lambda \left( \mS_{\vx^*}\mJ_H(\0) - \mJ_S(\0) \right) > 0$ is globally attractive in $B_y$. $\qedsymbol$
\end{theorem}\smallskip

\begin{figure}[!t]
    \centering
    \vspace{-5mm}
    \subfloat[For every point $p_k$, there is a point $(\vx_{rk}, \vy_{sk})$ starting from which, trajectories converge monotonically $(\leq_{K})$ to $(\vx^*,0)$.]{\includegraphics[scale=0.55]{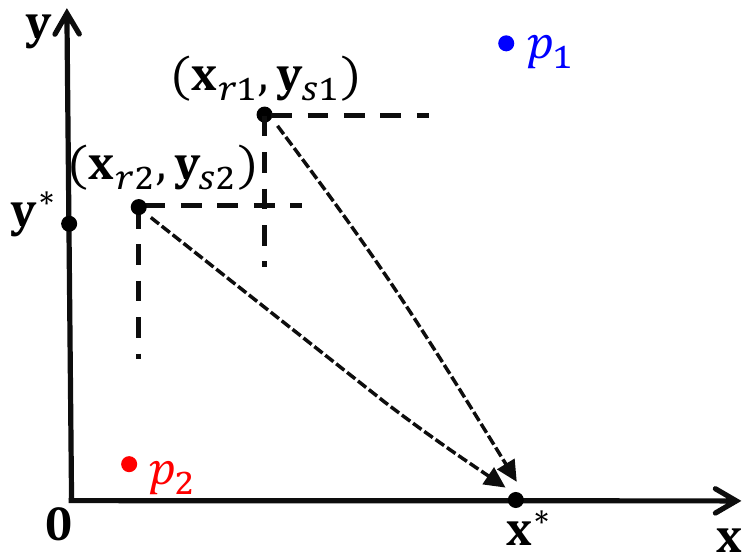}}
    \hfil
    \subfloat[Trajectories starting from $p_k$ eventually bounded by $(\vx_{rk}, \vy_{sk})$; monotonicity of the system gives convergence to $(\vx^*,0)$.]{\includegraphics[scale=0.55]{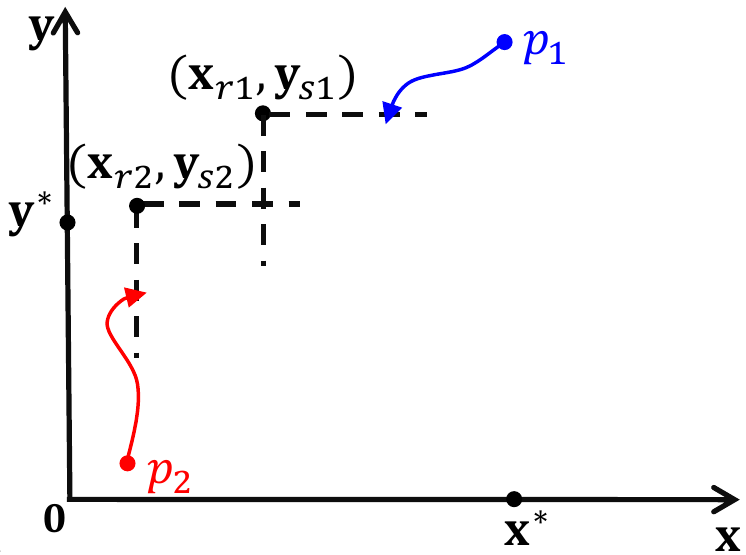}}
    \caption{Illustration of the convergence to $(\vx^*,0)$}\label{sketch}\vspace{-2mm}
\end{figure}

\begin{IEEEproof}[Sketch of the proof (convergence to $(\vx^*,\0)$)]
The idea behind the proof is illustrated in Figure \ref{sketch}. For every $(\vx,\vy)\!\in\!B_x$ (for example $p_1$ and $p_2$ in Figure \ref{sketch}), we construct a point $(\vx_r,\vy_s)$ which eventually bounds the trajectory starting from $(\vx,\vy)$; that is, we have $(\vx_r,\vy_s) \!\ll_K\! \phi_{t_1}(\vx,\vy) \!\leq_K\! (\vx^*,\0)$\footnote{$\phi_t(\vx,\vy)$ denotes the solution of \eqref{eq:biSIS} at $t\!\geq\!0$, with initial point $(\vx,\vy)$.} for some $t_1\!\geq\!0$. From the monotonicity shown in Proposition \ref{cooperative interor prop}, we have $\phi_t(\vx_r,\vy_s) \!\ll_K\! \phi_{t+t_1}(\vx,\vy) \!\leq_K\! (\vx^*,\0)$ for all time $t\!\geq\!0$. We prove that the trajectory starting from $(\vx_r,\vy_s)$ converges  to $(\vx^*,0)$ monotonically, with respect to the southeast cone-ordering (Figure \ref{sketch}(a)). Using this, we show the convergence of trajectories starting from $(\vx,\vy)$ via a sandwich argument (Figure \ref{sketch}(b)). See Appendix F in \cite{Appendices} for detailed proof.
\end{IEEEproof}\smallskip

\begin{figure*}[!htb]
    \centering
    \subfloat[Limitations of the literature.]{\includegraphics[scale=0.6]{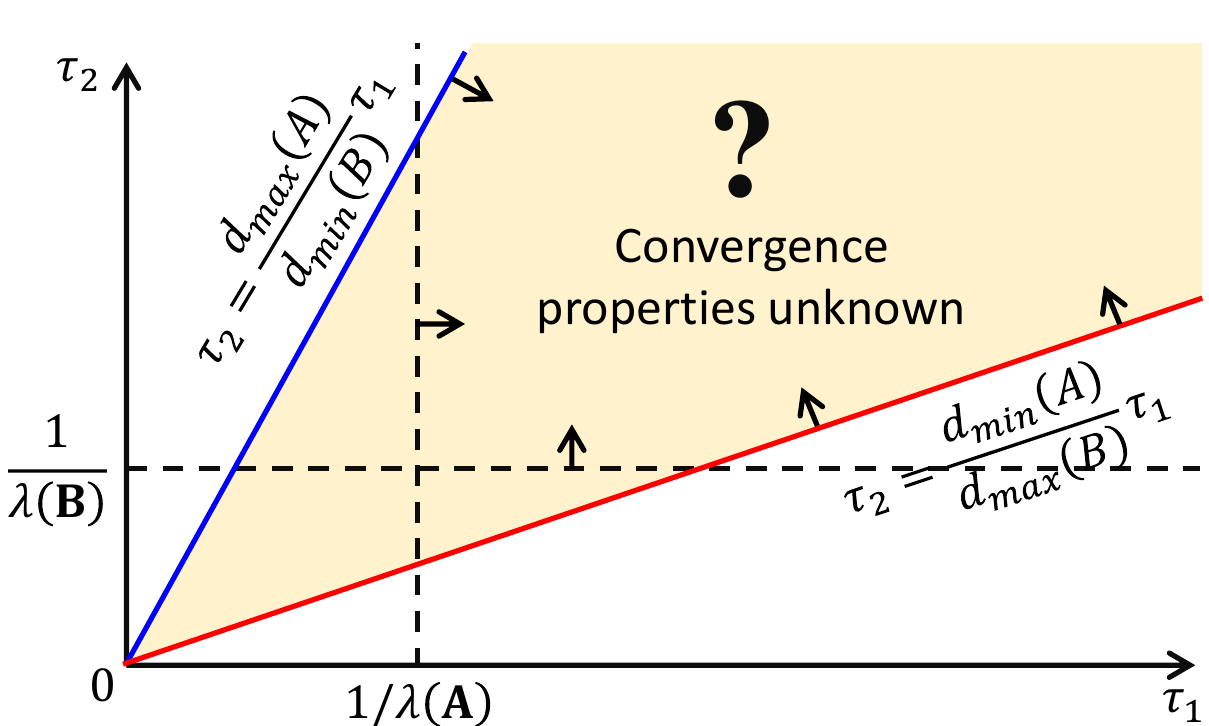}}
    \hfil
    \subfloat[Complete characterization of the convergence trichotomy.]{\includegraphics[scale=0.6]{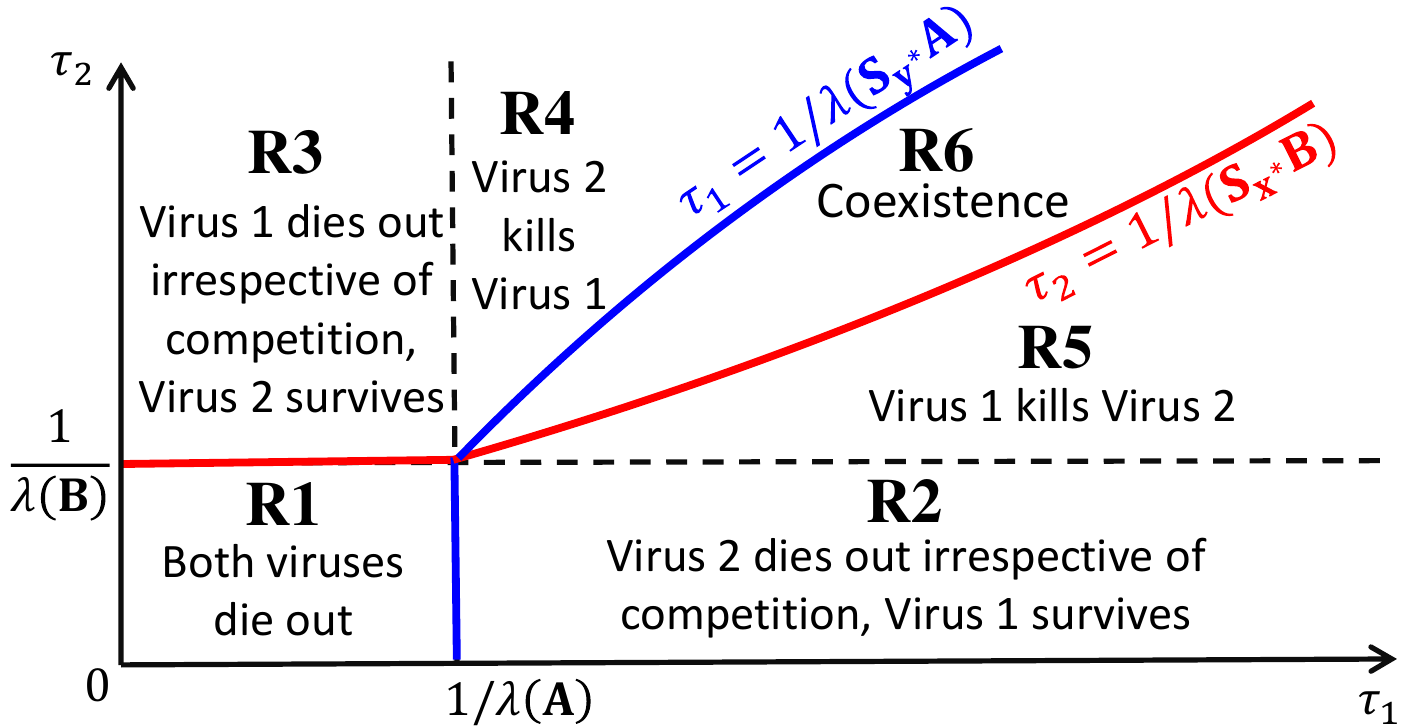}}
    \caption{Characterization of the parameter space}\label{all regions}\vspace{-2mm}
\end{figure*}

Finally, we give the necessary and sufficient conditions that guarantee the co-existence of the two viruses in the long run. Let $E$ denote the set of all fixed points of the system in \eqref{eq:biSIS-nonlinear}.
\smallskip
\begin{theorem}[Convergence to coexistence equilibria]\label{theorem coexistence}
  If $\lambda \left( \mS_{\vy^*}\mJ_G(\0) \!-\! \mJ_R(\0) \right) \!>\! 0$ and $\lambda \left( \mS_{\vx^*}\mJ_H(\0) \!-\! \mJ_S(\0) \right) \!>\! 0$, there exist fixed points of system \eqref{eq:biSIS-nonlinear} $(\hat \vx, \hat \vy) \!\gg\! (\0,\0)$ and $(\bar \vx,\bar \vy) \!\gg\! (\0,\0)$ such that
  \begin{equation*}
      (\0,\vy^*) \ll_K (\hat \vx,\hat \vy) \leq_K (\bar \vx,\bar \vy) \ll_K (\vx^*,\0),
  \end{equation*}
  \noindent with the possibility that $(\hat \vx, \hat \vy) = (\bar \vx,\bar \vy)$. All trajectories of system \eqref{eq:biSIS-nonlinear} starting from $B_x \cap B_y$ converge to the set of coexistence fixed points $S \triangleq \left\{ (\vx_e,\vy_e) \!\in\! E ~|~ (\hat \vx,\hat \vy) \!\leq_K\! (\vx_e,\vy_e) \!\leq_K\! (\bar \vx,\bar \vy)\right\}$.$\qedsymbol$
\end{theorem}\smallskip

The proof of Theorem \ref{theorem coexistence} follows similar arguments to that of the previous theorem, and is the first convergence result for coexistence fixed points in the competing SIS literature. Note that while we have convergence to `a' coexistence equilibrium, it may or may not be unique in the state space. The global convergence is therefore to the set of possible coexistence equilibria, and not necessarily a singular point. Thus, via Theorems \ref{theorem virus free}, \ref{theorem wta} and \ref{theorem coexistence} we cover all possible convergence scenarios of the bi-virus SIS system \eqref{eq:biSIS-nonlinear}, and successfully establish the complete theoretical characterization for the trichotomy of possible outcomes.

\section{Linear Infection and Recovery rates - Discussion and Comparison to Literature}\label{discussion linear}

We now take a look at the special case of the bi-virus epidemic model where infection and recovery rates scale linearly with the local infection probability. This is the most commonly analysed setting in literature \cite{liu2019analysis,R2,R3,R4,R5,R6}, and allows us to provide a comprehensive discussion on the related works. With the exception of \cite{R2}, a line of work seemingly developed concurrently to ours, we observe that most existing works only provide limited results regarding convergence to coexistence equilibria. In what follows, we provide corollaries of Theorems \ref{theorem virus free}, \ref{theorem wta} and \ref{theorem coexistence} which characterize convergence to the trichotomy of possible outcomes for the special case of linear infection and recovery rates. These results, along with Figure \ref{all regions}, are reproduced here as they originally were in our previous work \cite{vdoshi2021} which focused only on characterizing the convergence properties in the case of linear infection and recovery rates.

The model considered in this section is the bi-virus system \eqref{eq:biSIS} with \textit{homogeneous} infection and recovery rates\footnote{every infected node $i\in\cN$ infects its susceptible neighbor with the \textit{same} rate $\beta_1>0$ or $\beta_2>0$, and in turn recovers with the same rate $\delta_1>0$ or $\delta_2>0$, depending on whether it is infected by Virus 1 or 2 respectively.}. While at first this may seem too simplistic compared to the case of linear, \textit{heterogeneous} rates\footnote{The adjacency matrices $\mA$ and $\mB$ in \eqref{eq:biSIS} can be symmetric, irreducible, \textit{weighted}; with $a_{ij}, b_{ij} \geq 0$ (not necessarily $0/1$ valued) multiplied by $\beta_1$ and $\beta_2$ respectively, being the infection rates from node $j \to i$ for Viruses 1 and 2. Recovery rates can similarly be heterogenized as $\bfdelta_1 = [\delta_1^i]$ and $\bfdelta_2 = [\delta_2^i]$ for Viruses 1 and 2; written as recovery rate matrices $\text{diag}(\bfdelta_1)$ and $\text{diag}(\bfdelta_1)$, respectively.}, and even generic, non-linear rates analyzed in literature \cite{Santos2015,yang2017bi,liu2019analysis,R2,R3,R4,R5,R6}, the discussions in the `\textit{Comparison to existing ilterature}' subsection will still hold for these more general cases. We only stick to the bi-virus system with homogeneous rates as in \eqref{eq:biSIS} to be able to illustrate our results in the form of Figure \ref{all regions}; the axes capturing the parameters of the system. This enables us to better explain our contribution, using visual aids in the form of Figure \ref{all regions}, helping us compare our work with some of the existing literature more effectively, as opposed to presenting any other special case of the bi-virus model.


Consider the linear bi-virus system \eqref{eq:biSIS}. By setting $G(\vx) = \beta_1\mA\vx$, $R(\vx) = \delta_1 \vx$ and $H(\vy) = \beta_2 \mB \vy$, $S(\vy) = \delta_2 \vy$, we get
$$\mJ_G(\0) \!=\! \beta_1\mA, ~~\mJ_R(\0) \!=\! \delta_1 \eye,$$
and
$$\mJ_H(\0) \!=\! \beta_2\mB,~~ \mJ_S(\0) \!=\! \delta_2 \eye.$$
Defining $\tau_1 \triangleq \beta_1/\delta_1$, $\tau_2 = \triangle \beta_2/\delta_2$, and plugging in the above expressions for the Jacobians in Theorems \ref{theorem virus free} and \ref{theorem wta}, we have the following Corollaries.

\smallskip
\begin{corollary}\label{corollary virus free}
  If $\tau_1\lambda(\mA) \!\leq\! 1$ and $\tau_2\lambda(\mB) \!\leq\! 1$, trajectories of \eqref{eq:biSIS} starting from any point in $D$ converge to $(\0,\0)$.$\qedsymbol$
\end{corollary}\smallskip

\smallskip
\begin{corollary}\label{corollary wta}
  When $\tau_1\lambda(\mS_{\vy^*} \mA ) \!>\! 1$ and $\tau_2\lambda(\mS_{\vx^*} \mB) \!\leq\! 1$, $(\vx^*,\0)$ is globally attractive in $B_x$;\footnote{We consider $B_x$ as the global domain of attraction instead of $D$ because $\vx=0$ for all points in the set $D\setminus B_x$. Starting from such points the system is no longer a bi-virus epidemic, but a single-virus SIS system for Virus 2.} that is, every trajectory of system \eqref{eq:biSIS} starting from points in $B_x$ converges to $(\vx^*,\0)$.

  Similarly, when $\tau_1\lambda(\mS_{\vy^*} \mA ) \!\leq\! 1$ and $\tau_2\lambda(\mS_{\vx^*} \mB) \!>\! 1$, $(\0,\vy^*)$ is globally attractive in $B_y$. $\qedsymbol$
\end{corollary}\smallskip
From Corollary \ref{corollary wta}, we can deduce that the threshold values for $\tau_1$ and $\tau_2$ below which each of the viruses will die out are given by the equations $\tau_1\!=\!1/\lambda(\mS_{\vy^*} \mA )$ and $\tau_2\!=\!1/\lambda(\mS_{\vx^*} \mB )$, respectively. Figure \ref{all regions}(b) plots these threshold values for Virus 1 (in blue) and Virus 2 (in red) for varying values of $\tau_1$ and $\tau_2$, and partitions the entire parameter space into regions R1 -- R6 as shown. When $\tau_1 \!>\! 1/\lambda(\mA)$ and $\tau_2 \!>\!1/\lambda(\mB)$, for which values of $\tau_1,\tau_2$ do not lie in regions R1, R2 or R3, the blue curve lies above the red curve as in Figure \ref{all regions}(b). This was originally shown in \cite{sahneh2014competitive} by deducing that the ratio of slopes of the red and blue curves at point $(\tau_1,\tau_2) = \left(1/\lambda(\mA), 1/\lambda(\mB)\right)$ is less than one. This means there exist combinations of $\tau_1,\tau_2$ for which $\tau_1$ lies to the right of the blue curve ($\tau_1\lambda(\mS_{\vy^*} \mA ) \!>\! 1$), and $\tau_2$ lies above the red curve ($\tau_2\lambda(\mS_{\vx^*} \mB ) \!>\! 1$).\footnote{Note that $\tau_1\lambda(\mS_{\vy^*} \mA) \!\leq\! 1$ and $\tau_2\lambda(\mS_{\vx^*} \mB ) \!\leq\! 1$ is only possible in region R1, since it is the only region where $\tau_1$ can lie to the left of the blue curve, and $\tau_2$ can lie below the red curve. This effectively reduces the expressions to $\tau_1\lambda(\mA) \!\leq\! 1$ and $\tau_2\lambda(\mB ) \!\leq\! 1$, the conditions for convergence to the virus-free equilibrium as in Corollary \ref{corollary virus free}.} This corresponds to region R6 in Figure \ref{all regions}(b), and our final corollary (derived from Theorem \ref{theorem coexistence}) shows that for values of $\tau_1,\tau_2$ which lie in R6, we observe convergence to coexistence equilibria.

\smallskip
\begin{corollary}[Convergence to coexistence equilibria]\label{corollary coexistence}
  If $\tau_1\lambda(\mS_{\vy^*} \mA) \!>\! 1$ and $\tau_2\lambda(\mS_{\vx^*} \mB ) \!>\! 1$, there exist fixed points of system \eqref{eq:biSIS} $(\hat \vx, \hat \vy) \!\gg\! (\0,\0)$ and $(\bar \vx,\bar \vy) \!\gg\! (\0,\0)$ such that
  \begin{equation*}
      (\0,\vy^*) \ll_K (\hat \vx,\hat \vy) \leq_K (\bar \vx,\bar \vy) \ll_K (\vx^*,\0),
  \end{equation*}
  \noindent with the possibility that $(\hat \vx, \hat \vy) = (\bar \vx,\bar \vy)$. All trajectories of system \eqref{eq:biSIS} starting from $B_x \cap B_y$ converge to the set of coexistence fixed points $S \triangleq \left\{ (\vx_e,\vy_e) \!\in\! E ~|~ (\hat \vx,\hat \vy) \!\leq_K\! (\vx_e,\vy_e) \!\leq_K\! (\bar \vx,\bar \vy)\right\}$.$\qedsymbol$
\end{corollary}\smallskip
\begin{table*}[!t]
\centering \resizebox{12cm}{!}{%
\begin{tabular}{|c|l|l|l|l|}
\hline
\multicolumn{1}{|l|}{} & \multicolumn{1}{c|}{$g_i(\vx)$} & \multicolumn{1}{c|}{$h_i(\vy)$} & \multicolumn{1}{c|}{$r_i(\vx)$} & \multicolumn{1}{c|}{$s_i(\vy)$} \\ \hline
\textbf{CASE 1} & $\sum_{j} a_{ij} x_{j}$ & $ \sum_{j} b_{ij} y_{j}$ & $\delta_1 x_i$ & $\delta_2 y_i$ \\ \hline
\textbf{CASE 2} & $\sum_{j}a_{ij} \ln (1 + \alpha_1 x_j)$ & \multicolumn{1}{c|}{$\sum_{j}b_{ij} \ln (1 + \alpha_2 y_j)$} & $\delta_1 x_i$ & $\delta_2 y_i$ \\ \hline
\textbf{CASE 3} & \multicolumn{1}{c|}{$\sum_{j}a_{ij} \ln (1 + \alpha_1 x_j)$} & $\sum_{j}b_{ij} \ln (1 + \alpha_2 y_j)$ & $(1 + x_i)^2 - 1$ & $(1 + y_i)^2 - 1$ \\ \hline
\end{tabular}%
}
\caption{Summary of infection and recovery rate functions chosen.}\vspace{-4mm}
\label{tab:functions}
\end{table*}

\subsubsection*{\textbf{Comparison to existing literature}}
Now that we have established all our results, we briefly compare our work with results from \cite{yang2017bi,Santos2015}, which also talk about global convergence to single-virus equilibria. To this end, we first illustrate the limitations of the existing conditions for global convergence in \cite{yang2017bi,Santos2015} in Figure \ref{all regions}(a); and use Figure \ref{all regions}(b), where we provide complete characterization of the parameter space, to draw comparisons with our results. We then discuss the works \cite{R3,R6,R4,R5} which consider more general models where there can be more than two viruses, but present sharper results in the bi-virus setting. Finally, we will briefly comment on the finiteness of the coexistence equilibria, citing results from \cite{R2}.

When translated to the setting of linear infection and recovery rates as in \ref{eq:biSIS}, the result from \cite{Santos2015} says that when $\tau_1 d_{min}(\mA) \!>\! \tau_2 d_{max}(\mB)$, the Virus 2 is sure to die out (Virus 1 could persist or die out), and similarly when $\tau_1 d_{max}(\mA) \!<\! \tau_2 d_{min}(\mB)$, the Virus 1 is sure to die out. We illustrate these conditions in Figure \ref{all regions}(a), where Virus 1 (Virus 2) is sure to die out if parameters ($\tau_1,\tau_2$) lie above (below) the blue (red) line. Therefore, the entire yellow-shaded region in Figure \ref{all regions}(a), between the blue and red lines, is left uncharacterized in \cite{Santos2015}. When $\mA$ and $\mB$ are regular graphs with the same degree ($d_{min} \!=\! d_{max} \!=\! d$), the blue and red lines coincide, making coexistence infeasible. This is also mentioned in \cite{sahneh2014competitive} where they show that for regular graphs with same degree, the system behaves as if the two graphs were the same - rendering coexistence impossible (which is also in line with results in \cite{prakash2012winner}).
In contrast, the maximum degree of graphs can also be much larger than the minimum degree (e.g., power law graphs), causing the yellow-shaded space to become very large, possibly spanning almost the entire parameter space.

The main result in \cite{yang2017bi}, when similarly translated to our setting as above, says that when $\tau_1\lambda(\mA)\!>\!1$ and $\tau_2\lambda(\mB)\!\leq\!1$, Virus 1 survives and Virus 2 dies out. Similarly, when $\tau_2\lambda(\mB)\!>\!1$ and $\tau_1\lambda(\mA)\!\leq\!1$, Virus 2 survives and Virus 1 dies out. These correspond to regions R2 and R3 in Figure \ref{all regions}(b). However, their results do not cover the convergence properties for $\tau_1, \tau_2$ which lie in regions R4 -- R6. Our Theorems \ref{theorem wta} and \ref{theorem coexistence}, through their corresponding corollaries, do account for these values of $\tau_1, \tau_2$, and show convergence to $(\0,\vy^*)$, $(\vx^*,\0)$ or to a coexistence fixed point whenever they lie in regions R4, R5, or R6, respectively.

The works \cite{R4,R5} consider the bi-virus epidemic model with heterogeneous linear infection and recovery rates as a special case of their respective multi-virus models. Corollary 2 in \cite{R5}, a more general version of Theorem 5 in \cite{R4} which considers the case where $N=2$, establishes existence conditions for the coexistence equilibria. These conditions are identical to the ones emerging out of Theorem \ref{theorem coexistence} when applied to the bi-virus model considered therein (also identical to the conditions in Corollary \ref{corollary coexistence} for the special case of homogeneous, linear infection and recovery rates), and our result can therefore be considered as an extension of those in \cite{R4,R5}; providing \emph{convergence} results in addition to their existence results. Theorem 6 in \cite{R6} (Theorem 8 in \cite{R3}) is another interesting result concerning coexistence equilibria, where they show for the special case of viruses spreading over the same (possibly weighted) graph that the survival probability vectors of both the viruses are the same up to a constant multiple; that is, they are parallel.

The finiteness of the number of single-virus equilibria is evident from Theorem \ref{nonlinear sis-conditions}, which proves its uniqueness. However, Theorem \ref{theorem coexistence} and Corollary \ref{corollary coexistence} do not explicitly show that coexistence equilibria are finitely many, let alone uniqueness\footnote{In Section \ref{numerical results}, we show with the aid of simulation results that the coexistence equilibria are indeed not unique in general.}. For linear, heterogeneous infection and recovery rates, Theorem 3.6 in \cite{R2} uses novel techniques from algebraic geometry to prove that the coexistence equilibria are finitely many for all possible values of infection and recovery rates that do not lie in an algebraic set of measure zero. However, this remains an open problem for general, non-linear infection and recovery rate functions satisfying (A1)--(A5).

In summary, without our Theorems \ref{theorem wta} and \ref{theorem coexistence}, convergence results from literature fail to characterize a sizeable portion of the parameter space as shown in Figure \ref{all regions}(a) by the `\textbf{?}' region (part of the shaded region surrounded by the arrows). The parameters leading to coexistence are entirely contained in this region as well - explaining the dearth of convergence results for such equilibria in the existing literature.

\section{Numerical Results}\label{numerical results}

In this section, we present simulation results to support our theoretical findings for the bi-virus SIS model for combinations of non-linear as well as linear infection and recovery rates. To this end, we consider an undirected, connected graph (103 nodes, 239 edges), called Autonomous System (AS-733), from the SNAP repository~\cite{snapnets}. For both the linear and non-linear bi-virus model, we generate an additional graph, overlaid on the same set of nodes, by modifying the original graph (AS-733-A with $\lambda(\mA) \!=\! 12.16$), removing and adding edges while ensuring connectivity between the nodes. The new additional graph, AS-733-B, has 741 edges with $\lambda(\mB) \!=\! 15.53$. Note that since our theoretical results hold for any general graphs, we only use this set as example graphs to numerically demonstrate the convergence properties. Similar numerical results can indeed be obtained for any other networks (such as social networks).

We test the convergence dynamics of the bi-virus model over a range of combinations of linear and non-linear infection and recovery rates. To this end, we consider three different bi-virus models, and Table \ref{tab:functions} summarizes the three cases with the corresponding infection and recovery rate functions as shown. Note that for non-linear infection and recovery rates, we consider the logarithmic and polynomial functions briefly mentioned in Section \ref{nonlinear epidemic models}, to ensure that our three cases satisfy assumptions (A1)--(A5).

\begin{figure*}[!t]
    \centering
    \captionsetup{justification=centering}
    \subfloat[$\lambda(U)>0$, $\lambda(V)<0$; Virus 1 survives]
    {\includegraphics[scale=0.28]{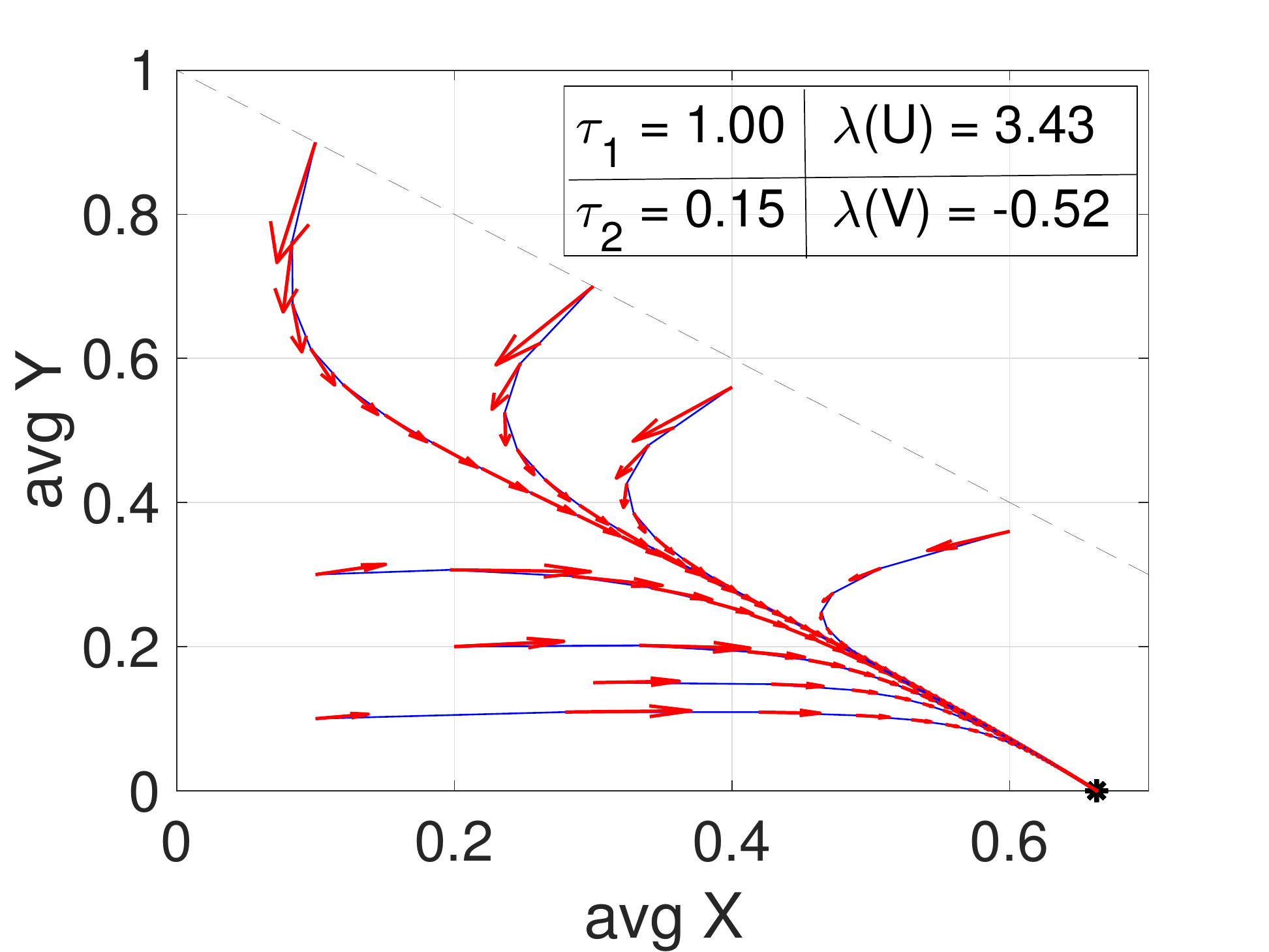}}
    \hfil
    \subfloat[$\lambda(U)<0$, $\lambda(V)>0$; Virus 2 survives]{\includegraphics[scale=0.28]{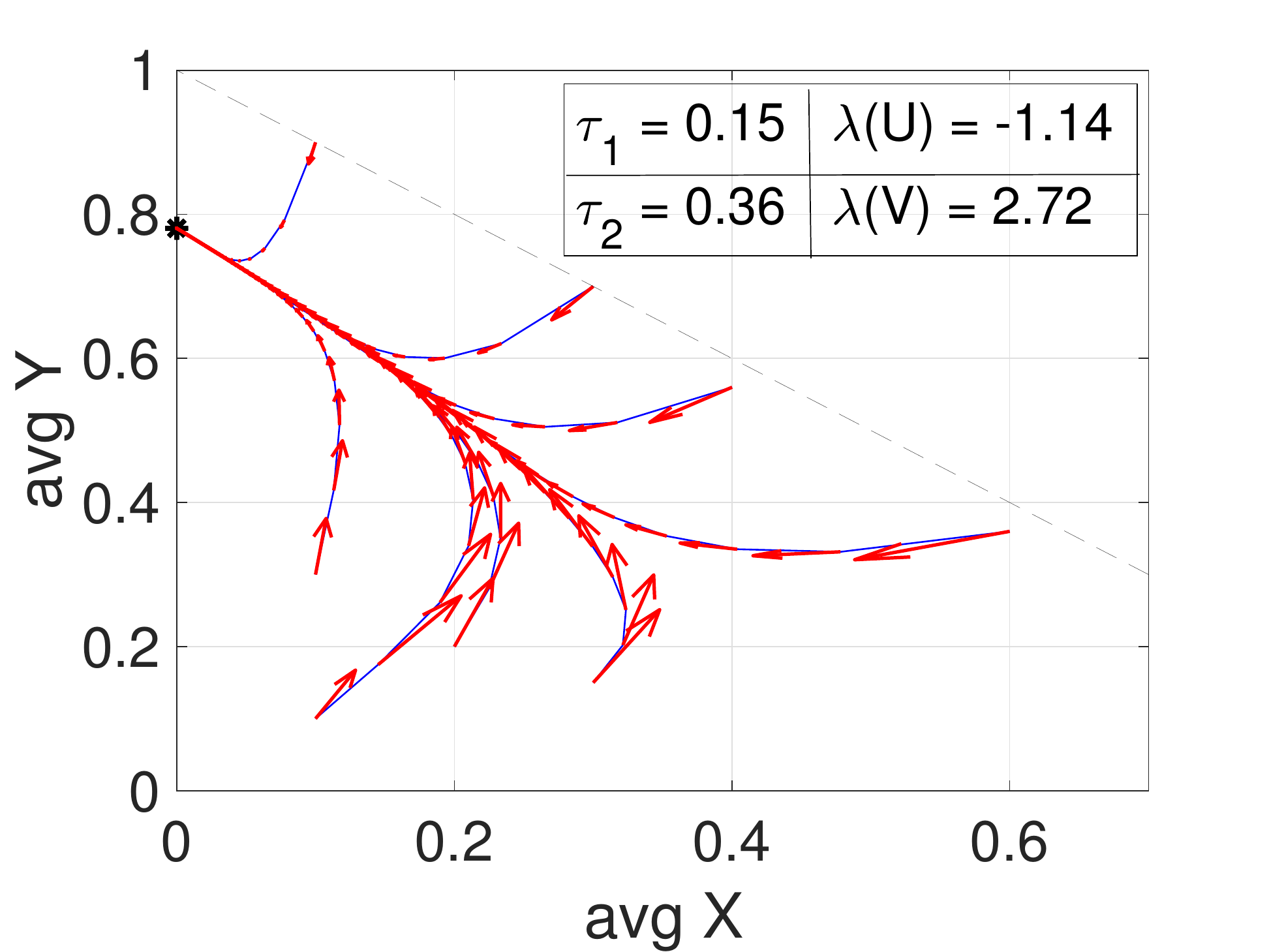}}%
    \hfil
    \subfloat[$\lambda(U)>0$, $\lambda(V)>0$; Both coexist]{\includegraphics[scale=0.28]{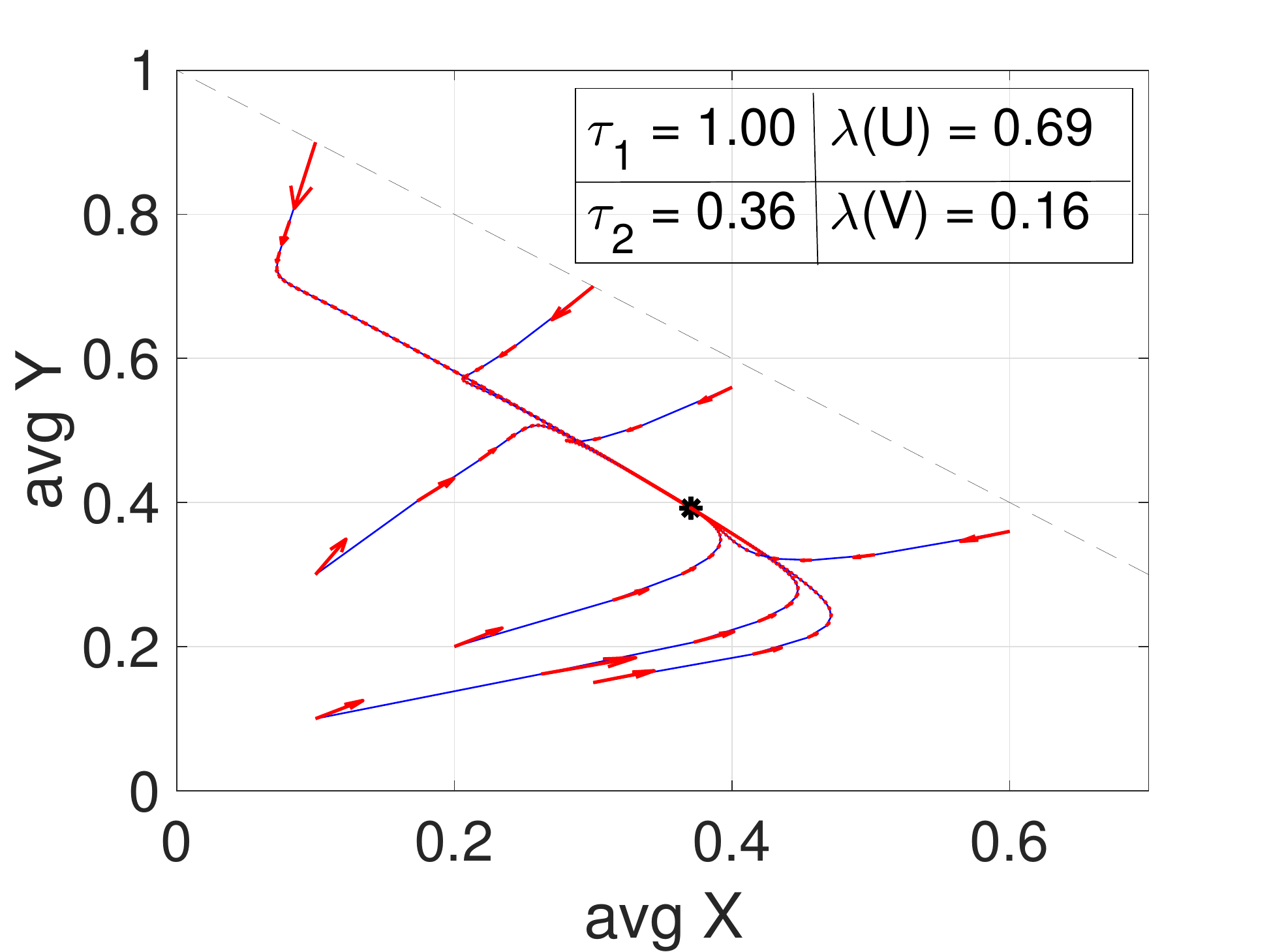}}
    \caption{Phase plots for a system with linear infection and recovery rates (CASE 1) on the AS-733 graph.}\vspace{-2mm}
    \label{fig:linear}
\end{figure*}

\begin{figure*}[!t]\vspace{-4mm}
    \centering
    \captionsetup{justification=centering}
    \subfloat[$\lambda(U)>0$, $\lambda(V)<0$; Virus 1 survives]
    {\includegraphics[scale=0.28]{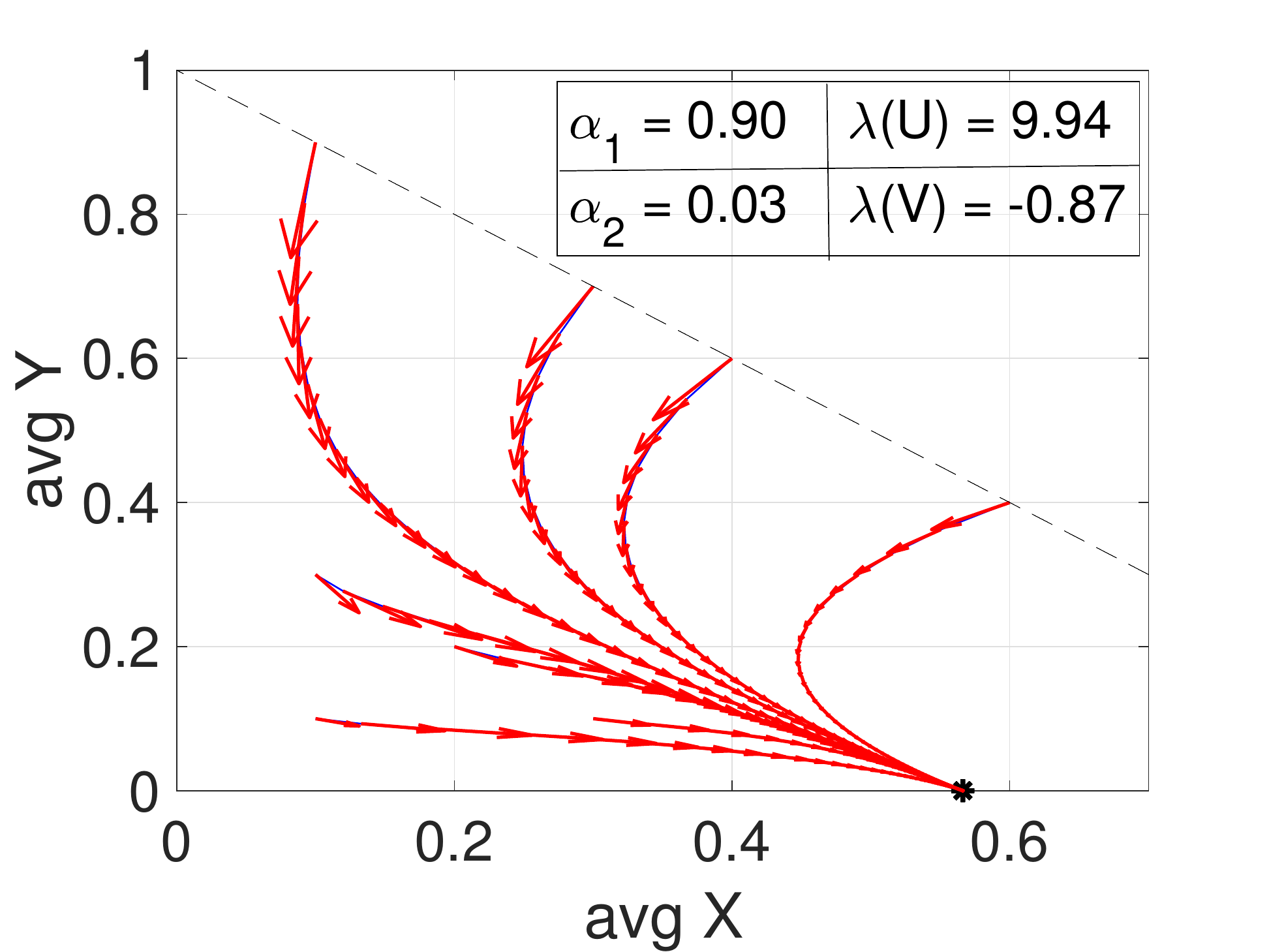}}
    \hfil
    \subfloat[$\lambda(U)<0$, $\lambda(V)>0$; Virus 2 survives]{\includegraphics[scale=0.28]{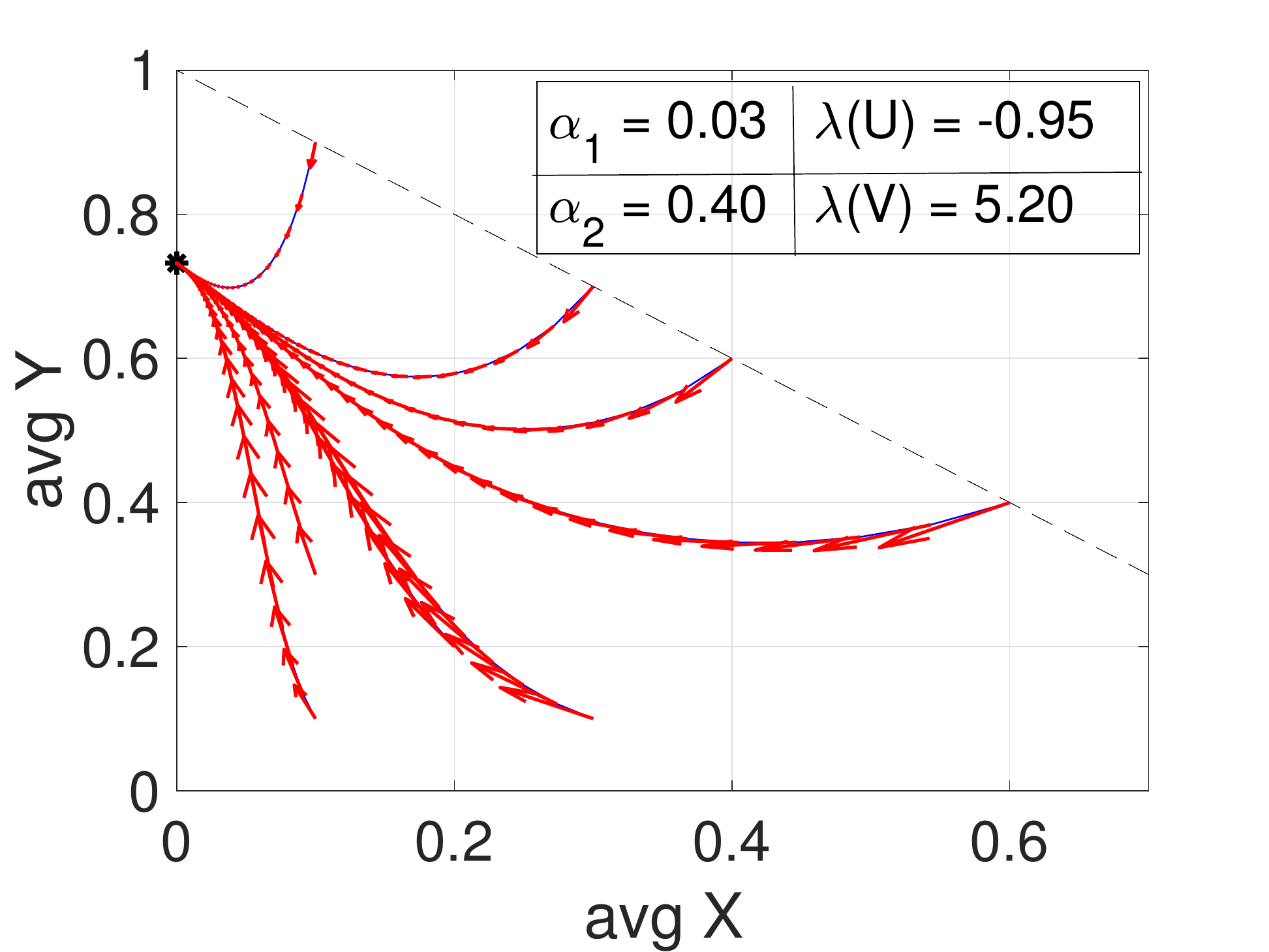}}%
    \hfil
    \subfloat[$\lambda(U)>0$, $\lambda(V)>0$; Both coexist]{\includegraphics[scale=0.28]{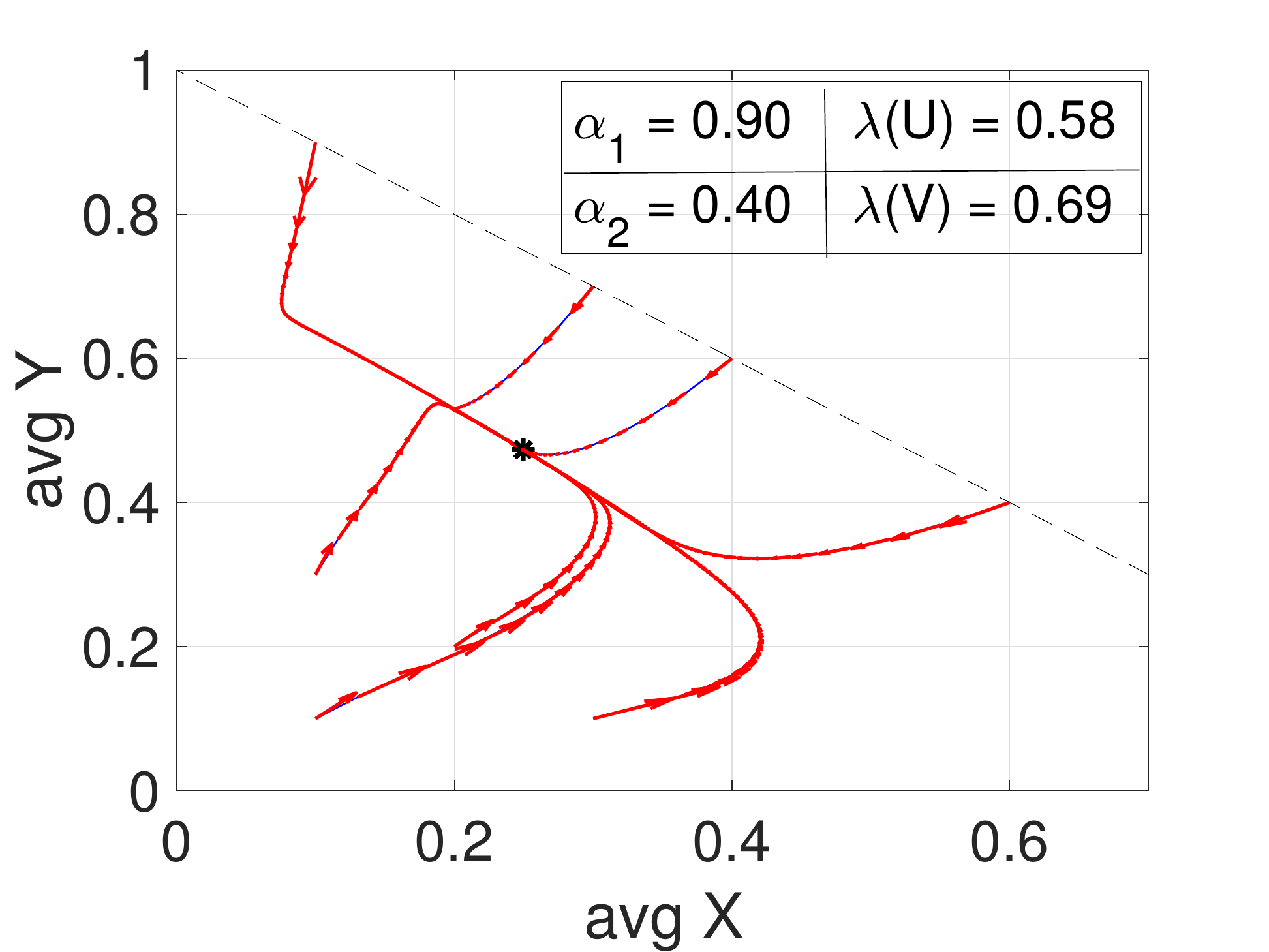}}
    \caption{Phase plots for a system with non-linear infection and linear recovery rates (CASE 2) on the AS-733 graph.}\vspace{-2mm}
    \label{fig:case2}
\end{figure*}

\begin{figure*}[!t]\vspace{-4mm}
    \centering
    \captionsetup{justification=centering}
    \subfloat[$\lambda(U)>0$, $\lambda(V)<0$; Virus 1 survives]
    {\includegraphics[scale=0.28]{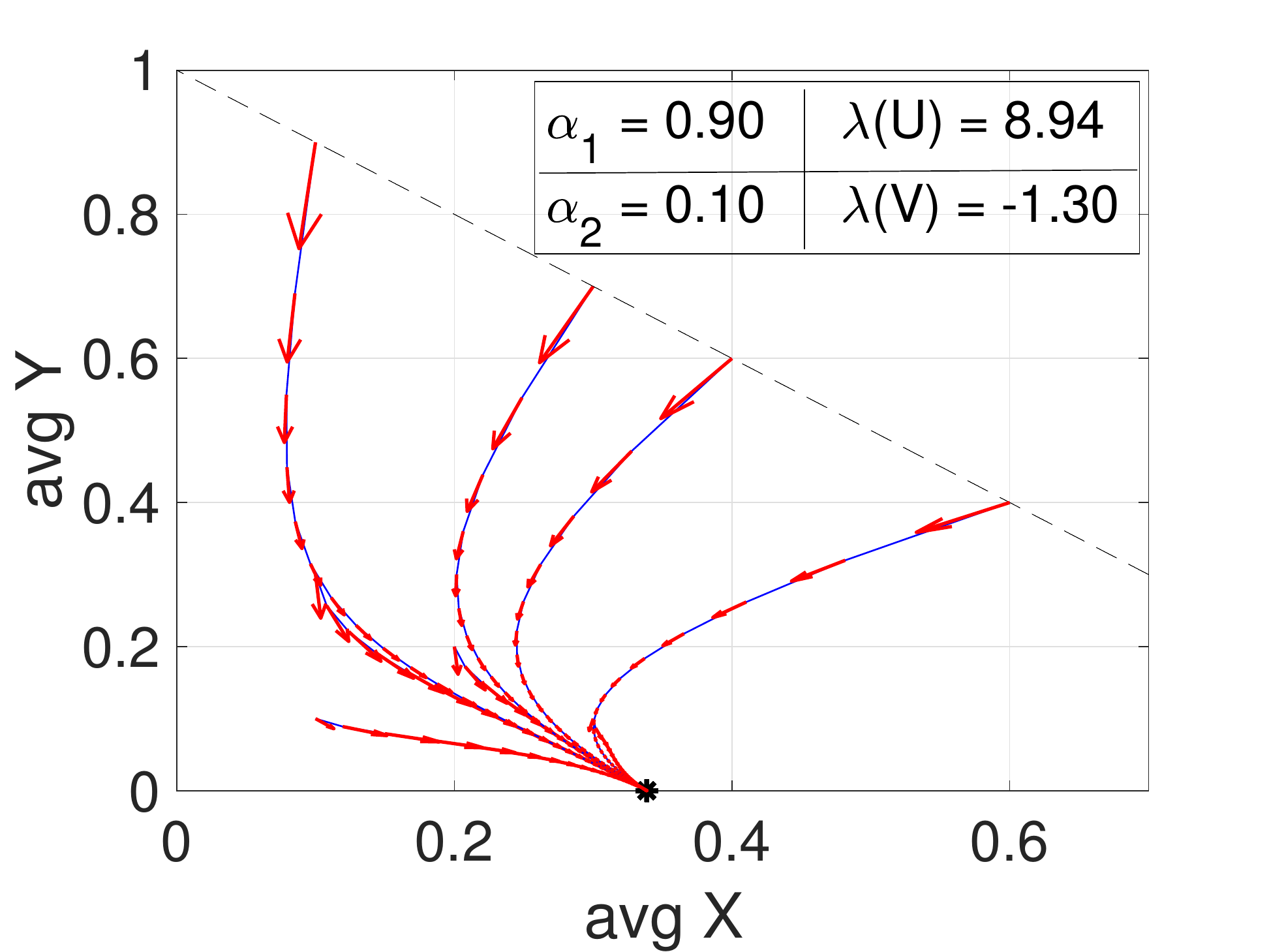}}
    \hfil
    \subfloat[$\lambda(U)<0$, $\lambda(V)>0$; Virus 2 survives]{\includegraphics[scale=0.28]{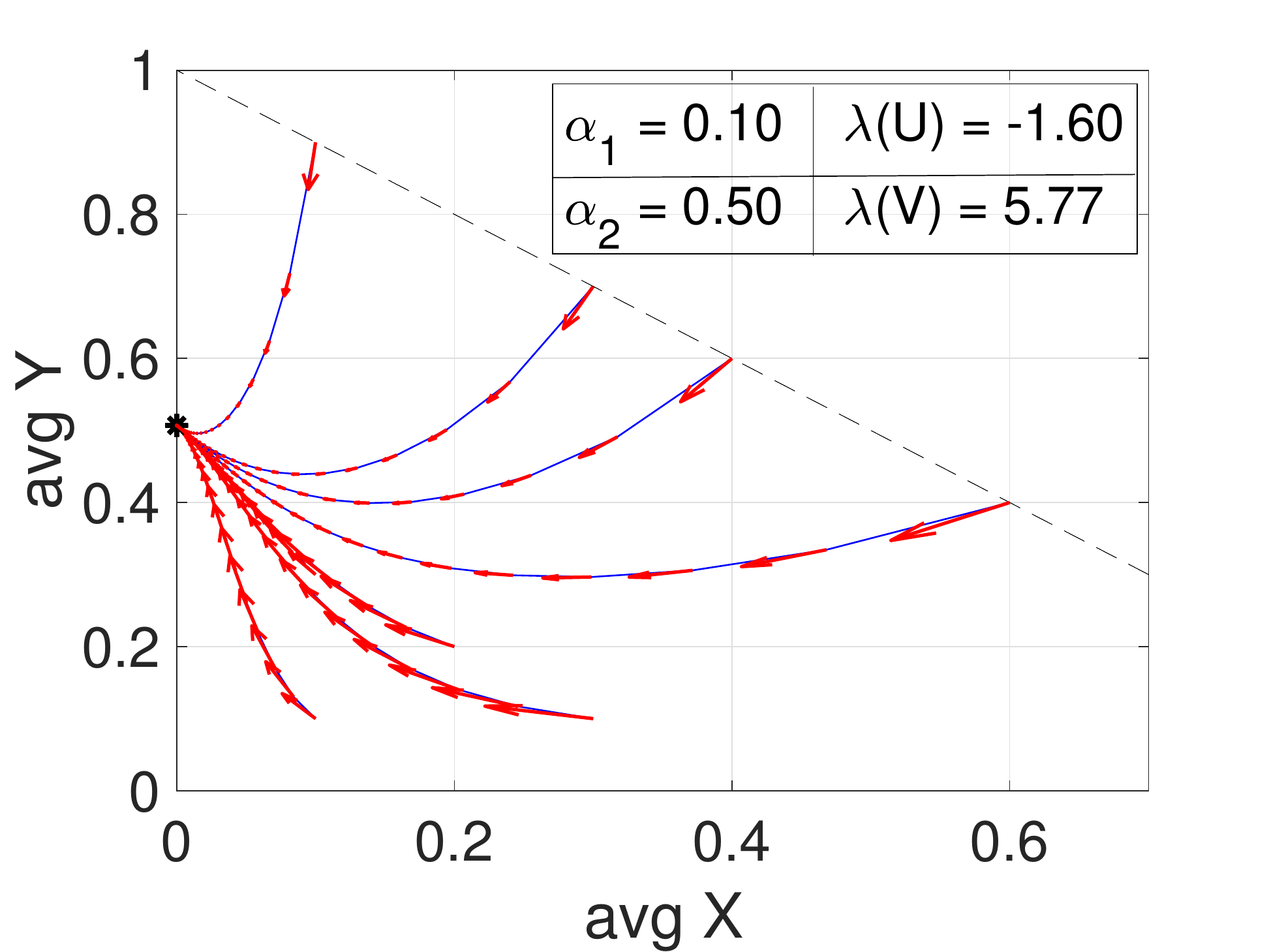}}%
    \hfil
    \subfloat[$\lambda(U)>0$, $\lambda(V)>0$; Both coexist]{\includegraphics[scale=0.28]{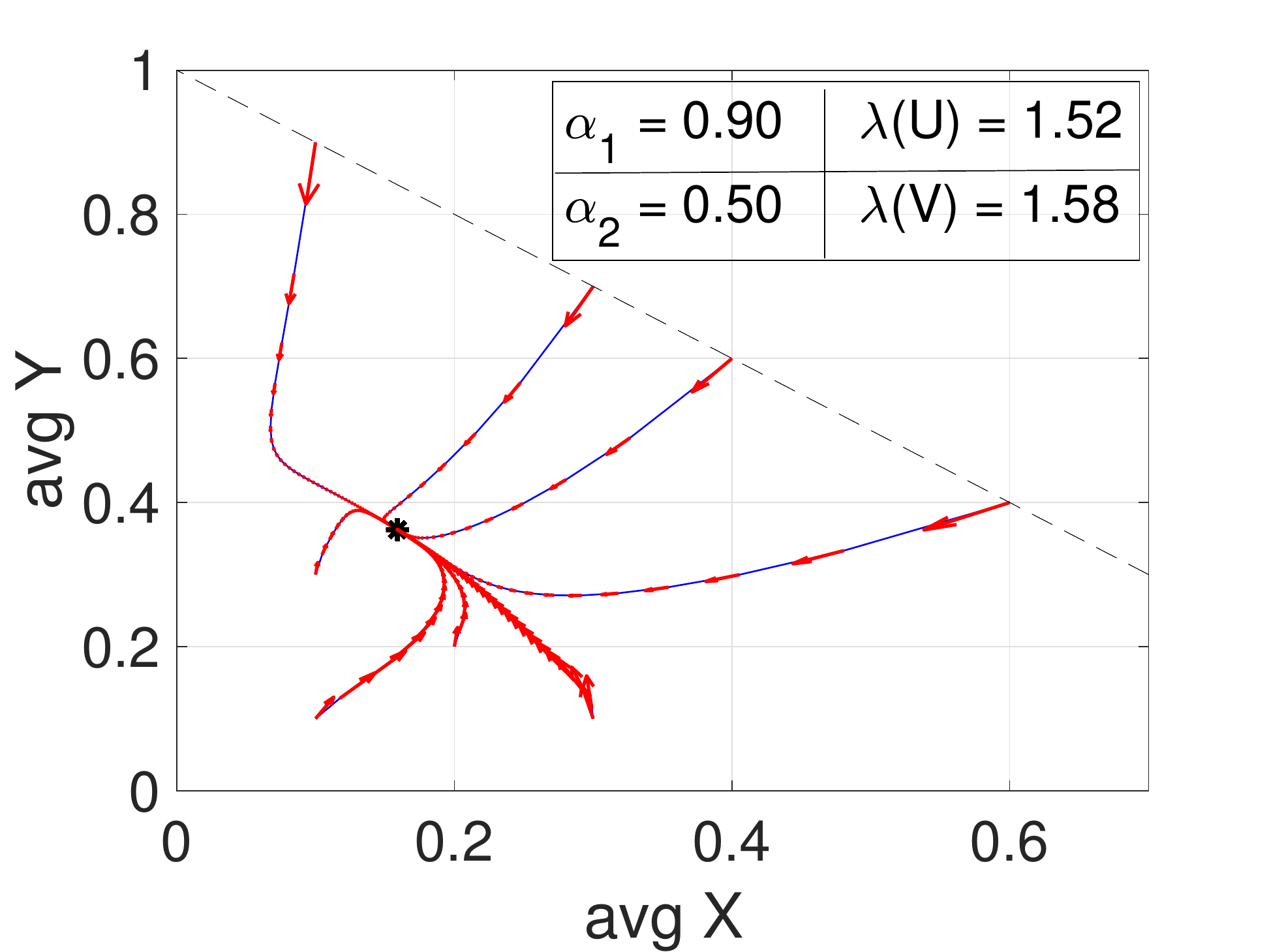}}
    \caption{Phase plots for a system with non-linear infection and recovery rates (CASE 3) on the AS-733 graph.}\vspace{-2mm}
    \label{fig:case3}
\end{figure*}

For each of the three cases, we construct combinations of parameters ($\tau_1$ or $\tau_2$ for linear rates, and $\alpha_1$ or $\alpha_2$ for non-linear rates), to develop three convergence scenarios, that satisfy the assumptions of Theorems \ref{theorem wta} and \ref{theorem coexistence}. These three scenarios correspond to global convergence of the bi-virus system to fixed points where (a) Virus 1 is the surviving epidemic (which spreads on graph AS-733-A), (b) Virus 2 is the surviving epidemic (which spreads on graph AS-733-B), (c) both viruses coexist, (where Virus 1 spreads on graph AS-733-A and Virus 2 on AS-733-B). Parameters corresponding to these three scenarios are provided in the table inset in Figures~\ref{fig:linear}--\ref{fig:case3}(a)--(c) corresponding to the three cases.

To visualize our system in two dimensions, we use $avg X \!\triangleq\! (1/N)\sum_{i \in \cN} x_i$ on the x-axis, and $avg Y \!\triangleq\! (1/N)\sum_{i \in \cN} y_i$ on the y-axis. We plot trajectories of the bi-virus system starting from different initial points in the state space $D$ to observe their convergence, with red arrows representing the trajectories' direction of movement at various time intervals. Here, the state space $D$ is the region that lies below the dotted-line (for example, in Figure~\ref{fig:linear}), ensuring $x_i + y_i \!\leq\! 1$ for all $i \in \cN$, for every initial point. To ensure that the convergences observed in our phase plots match the conditions laid out in Theorems~\ref{theorem wta} and~\ref{theorem coexistence}, we track the eigenvalues $\lambda(\mU) \triangleq \lambda(\mS_{\vy^*}\mJ_G(0) - \mJ_R(0))$ and $\lambda(\mV) \triangleq \lambda(\mS_{\vx^*}\mJ_H(0) - \mJ_S(0))$. $\lambda(\mU)$ ($\lambda(\mV)$) being positive or negative corresponds to Virus 1 (Virus 2) surviving or dying out, respectively.

\begin{figure}[!h]\vspace{-4mm}
    \centering
    {\includegraphics[scale=0.35]{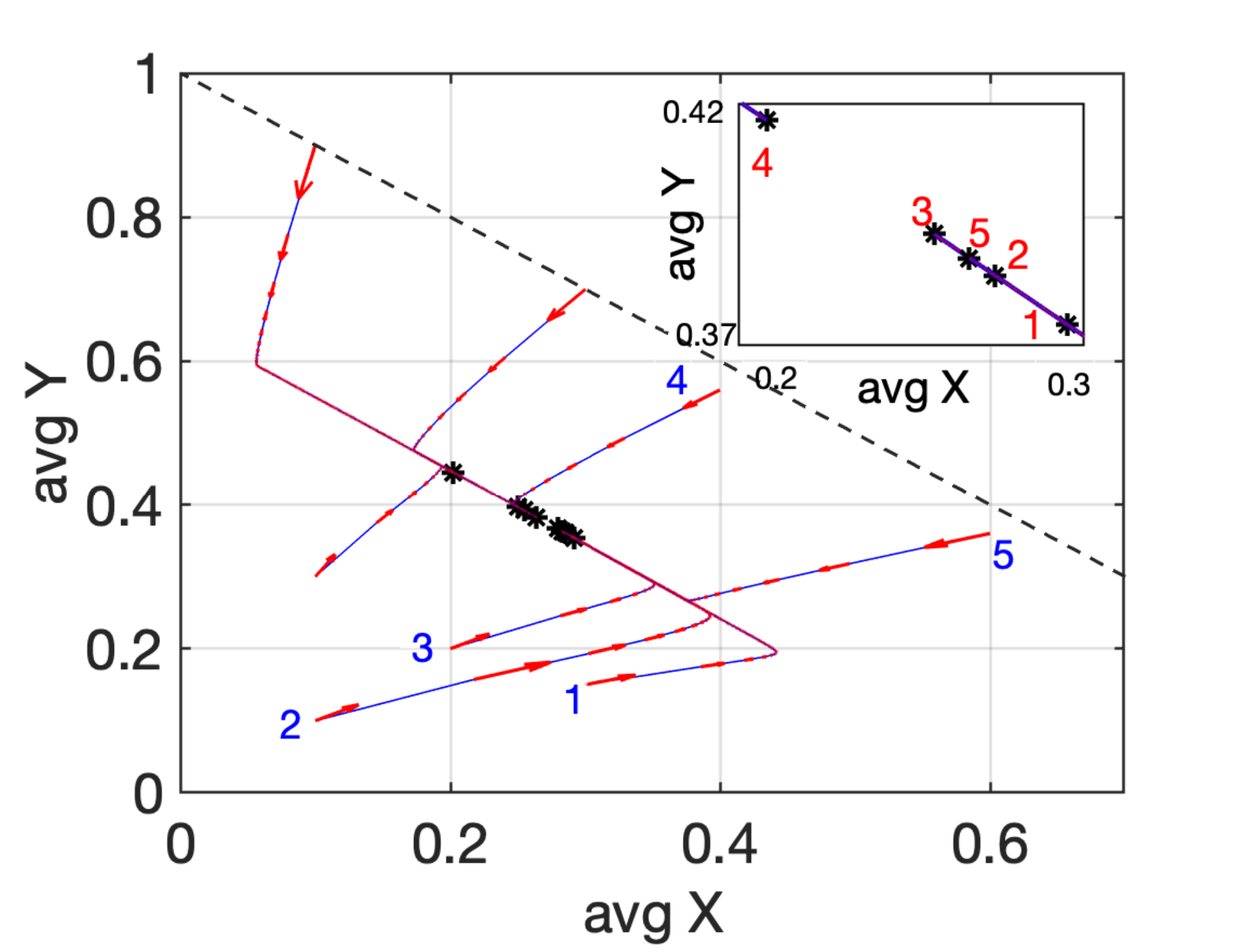}}
    \caption{Coexistence condition with Multiple equilibrium points}\vspace{-2mm}
    \label{fig:multiple}
\end{figure}

In Figures~\ref{fig:linear}--\ref{fig:case3}(a)--(c), we show numerical results for the three cases, respectively. Figures~\ref{fig:linear}--\ref{fig:case3}(a) and~\ref{fig:linear}--\ref{fig:case3}(b) show convergence to the two different single-virus equilibria, where the parameters therein satisfy the two set of conditions as in Theorem~\ref{theorem wta}. Figures~\ref{fig:linear}--\ref{fig:case3}(c) show convergence to the coexistence equilibria, which also satisfies the coexistence conditions as outlined in Theorem~\ref{theorem coexistence}. We observe a unique
coexistence equilibrium when the viruses are competing over graphs AS-733-A and AS-733-B, for which the eigenvalues
$\lambda(\mA)$ and $\lambda(\mB)$ are significantly different. Interestingly, we also observe multiple coexistence equilibria as shown in Figure~\ref{fig:multiple}. We obtain this result by creating another additional graph by modifying the original graph AS-733-A such that the eigenvalue of this new graph is as close to the original one where this new graph AS-733-C has 259 edges with $\lambda(\mC)\!\! =\!\! 12.26$. The `upper left' and `lower right' coexistence fixed points characterize the set $S$ of all such equilibria, as in Theorem~\ref{theorem coexistence}. This can be seen more closely in the inset in Figure~\ref{fig:multiple}, where the number beside each fixed point (in red) corresponds to the different initial starting points (in blue) of the trajectories. Thus, convergence to set $S$ occurs globally over the state space, but exactly which coexistence fixed point the system converges to is dependent on the initial point. We are thus able to observe all possible convergence scenarios from Section~\ref{convergence and coexistence}, including multiple coexistence equilibria.

\section{Concluding Remarks} \label{conclusion}

By utilizing the techniques from Monotone Dynamical Systems (MDS), in this paper, we show that a generic bi-virus epidemic model with non-linear infection and recovery rates is monotone with respect to a specially constructed partial ordering. This monotonicity allows us to give necessary and sufficient conditions on the non-linear infection and recovery rates, and thus completely characterize the entire parameter space of the bi-virus system, a contrast to the usual Lyapunov based approach. We bridge the gap between linear stability properties and global convergence results (or lack thereof) for the bi-virus model with non-linear rates (including the special case with linear rates) in the literature, and succeed in providing a complete characterization of the trichotomy of possible outcomes for such competing epidemics - a well known open problem. Our results demonstrate how powerful these alternative proving techniques can be, compared to classical Lyapunov approaches; and we note that it may be worth exploring such monotonicity properties in other dynamics on graphs as well, where competition is a general theme. Additionally, establishing a rigorous relationship between the SIS ODE models with non-linear rates as studied in this paper, and the correct probabilistic dynamics describing these non-linear rates, is of interest in order to complete the theoretical pictures for SIS models with non-linear rates.

\bibliographystyle{IEEEtran}
\bibliography{sis}

\begin{thebibliography}{10}
\providecommand{\url}[1]{#1}
\csname url@samestyle\endcsname
\providecommand{\newblock}{\relax}
\providecommand{\bibinfo}[2]{#2}
\providecommand{\BIBentrySTDinterwordspacing}{\spaceskip=0pt\relax}
\providecommand{\BIBentryALTinterwordstretchfactor}{4}
\providecommand{\BIBentryALTinterwordspacing}{\spaceskip=\fontdimen2\font plus
\BIBentryALTinterwordstretchfactor\fontdimen3\font minus
  \fontdimen4\font\relax}
\providecommand{\BIBforeignlanguage}[2]{{%
\expandafter\ifx\csname l@#1\endcsname\relax
\typeout{** WARNING: IEEEtran.bst: No hyphenation pattern has been}%
\typeout{** loaded for the language `#1'. Using the pattern for}%
\typeout{** the default language instead.}%
\else
\language=\csname l@#1\endcsname
\fi
#2}}
\providecommand{\BIBdecl}{\relax}
\BIBdecl

\bibitem{vdoshi2021}
V.~Doshi, S.~Mallick, and D.~Y. Eun, ``{C}ompeting {E}pidemics on {G}raphs -
  {G}lobal {C}onvergence and {C}oexistence,'' in \emph{IEEE INFOCOM}, 2021.

\bibitem{Yorke1976}
A.~Lajmanovich and J.~A. Yorke, ``A deterministic model for gonorrhea in a
  nonhomogeneous population,'' \emph{Mathematical Biosciences}, vol.~28, no.~3,
  pp. 221 -- 236, 1976.

\bibitem{hethcote2000mathematics}
H.~W. Hethcote, ``The mathematics of infectious diseases,'' \emph{SIAM Review},
  vol.~42, no.~4, pp. 599--653, 2000.

\bibitem{garetto2003modeling}
M.~Garetto, W.~Gong, and D.~Towsley, ``Modeling malware spreading dynamics,''
  in \emph{IEEE INFOCOM}, San Francisco, CA, 2003.

\bibitem{yang2013epidemics}
L.-X. Yang, X.~Yang, J.~Liu, Q.~Zhu, and C.~Gan, ``Epidemics of computer
  viruses: a complex-network approach,'' \emph{Applied Mathematics and
  Computation}, vol. 219, no.~16, pp. 8705--8717, 2013.

\bibitem{hosseini2016model}
S.~Hosseini and M.~A. Azgomi, ``A model for malware propagation in scale-free
  networks based on rumor spreading process,'' \emph{Computer Networks}, vol.
  108, pp. 97--107, 2016.

\bibitem{apt2011diffusion}
K.~R. Apt and E.~Markakis, ``Diffusion in social networks with competing
  products,'' in \emph{International Symposium on Algorithmic Game Theory},
  2011.

\bibitem{prakash2012winner}
B.~A. Prakash, A.~Beutel, R.~Rosenfeld, and C.~Faloutsos, ``Winner takes all:
  competing viruses or ideas on fair-play networks,'' in \emph{ACM World Wide
  Web}, 2012.

\bibitem{ruf2017dynamics}
S.~F. Ruf, K.~Paarporn, P.~E. Pare, and M.~Egerstedt, ``Dynamics of
  opinion-dependent product spread,'' in \emph{IEEE Conference on Decision and
  Control}, Melbourne, Australia, 2017.

\bibitem{trpevski2010model}
D.~Trpevski, W.~K. Tang, and L.~Kocarev, ``Model for rumor spreading over
  networks,'' \emph{Physical Review E}, vol.~81, no.~5, p. 056102, 2010.

\bibitem{zhao2013sir}
L.~Zhao, H.~Cui, X.~Qiu, X.~Wang, and J.~Wang, ``{SIR} rumor spreading model in
  the new media age,'' \emph{Physica A: Statistical Mechanics and its
  Applications}, vol. 392, no.~4, pp. 995--1003, 2013.

\bibitem{lin2018opinion}
X.~Lin, Q.~Jiao, and L.~Wang, ``Opinion propagation over signed networks:
  Models and convergence analysis,'' \emph{IEEE Transactions on Automatic
  Control}, vol.~64, no.~8, pp. 3431--3438, 2018.

\bibitem{Shroff2019}
I.~Koprulu, Y.~Kim, and N.~B. Shroff, ``Battle of opinions over evolving social
  networks,'' \emph{IEEE/ACM Transactions on Networking}, vol.~27, no.~2, pp.
  532--545, 2019.

\bibitem{shakkottaiINFOCOM2014}
S.~Banerjee, A.~Chatterjee, and S.~Shakkottai, ``Epidemic thresholds with
  external agents,'' in \emph{IEEE INFOCOM}, Toronto, ON, 2014.

\bibitem{ganesh2005effect}
A.~Ganesh, L.~Massoulie, and D.~Towsley, ``The effect of network topology on
  the spread of epidemics,'' in \emph{IEEE INFOCOM}, Miami, FL, 2005.

\bibitem{draief_massoulie_2009}
M.~Draief and L.~Massoulié, \emph{Epidemics and Rumours in Complex Networks},
  1st~ed.\hskip 1em plus 0.5em minus 0.4em\relax Cambridge University Press,
  2010.

\bibitem{Mieghem2013}
F.~Darabi~Sahneh, C.~Scoglio, and P.~Van~Mieghem, ``Generalized epidemic
  mean-field model for spreading processes over multilayer complex networks,''
  \emph{IEEE/ACM Transactions on Networking}, vol.~21, no.~5, pp. 1609--1620,
  2013.

\bibitem{sahneh2014competitive}
F.~D. Sahneh and C.~Scoglio, ``Competitive epidemic spreading over arbitrary
  multilayer networks,'' \emph{Physical Review E}, vol.~89, no.~6, p. 062817,
  2014.

\bibitem{Santos2015}
A.~{Santos}, J.~M.~F. {Moura}, and J.~M.~F. {Xavier}, ``Bi-virus {SIS}
  epidemics over networks: Qualitative analysis,'' \emph{IEEE Transactions on
  Network Science and Engineering}, vol.~2, no.~1, pp. 17--29, 2015.

\bibitem{yang2017bi}
L.-X. Yang, X.~Yang, and Y.~Y. Tang, ``A bi-virus competing spreading model
  with generic infection rates,'' \emph{IEEE Transactions on Network Science
  and Engineering}, vol.~5, no.~1, pp. 2--13, 2017.

\bibitem{liu2019analysis}
J.~Liu, P.~E. Par{\'e}, A.~Nedich, C.~Y. Tang, C.~L. Beck, and T.~Basar,
  ``Analysis and control of a continuous-time bi-virus model,'' \emph{IEEE
  Transactions on Automatic Control}, 2019.

\bibitem{n-intertwined}
P.~Van~Mieghem, ``The n-intertwined {SIS} epidemic network model,''
  \emph{Computing}, vol.~93, no. 2–4, p. 147–169, 2011.

\bibitem{omic2009epidemic}
J.~Omic and P.~Van~Mieghem, ``Epidemic spreading in networks—variance of the
  number of infected nodes,'' \emph{Delft University of Technology, Report},
  2009.

\bibitem{van2009virus}
P.~Van~Mieghem, J.~Omic, and R.~Kooij, ``Virus spread in networks,''
  \emph{IEEE/ACM Transactions on Networking}, vol.~17, no.~1, pp. 1--14, 2009.

\bibitem{gray2011stochastic}
A.~Gray, D.~Greenhalgh, L.~Hu, X.~Mao, and J.~Pan, ``A stochastic differential
  equation {SIS} epidemic model,'' \emph{SIAM Journal on Applied Mathematics},
  vol.~71, no.~3, pp. 876--902, 2011.

\bibitem{li2012susceptible}
C.~Li, R.~van~de Bovenkamp, and P.~Van~Mieghem,
  ``Susceptible-infected-susceptible model: A comparison of n-intertwined and
  heterogeneous mean-field approximations,'' \emph{Physical Review E}, vol.~86,
  no.~2, p. 026116, 2012.

\bibitem{wang2012global}
Y.~Wang, Z.~Jin, Z.~Yang, Z.-K. Zhang, T.~Zhou, and G.-Q. Sun, ``Global
  analysis of an {SIS} model with an infective vector on complex networks,''
  \emph{Nonlinear Analysis: Real World Applications}, vol.~13, no.~2, pp.
  543--557, 2012.

\bibitem{guo2013epidemic}
D.~Guo, S.~Trajanovski, R.~van~de Bovenkamp, H.~Wang, and P.~Van~Mieghem,
  ``Epidemic threshold and topological structure of
  susceptible-infectious-susceptible epidemics in adaptive networks,''
  \emph{Physical Review E}, vol.~88, no.~4, p. 042802, 2013.

\bibitem{benaim1999differential}
M.~Bena\"{i}m and M.~W. Hirsch, ``Differential and stochastic epidemic
  models,'' \emph{Fields Institute communications}, vol.~21, pp. 31--44, 1999.

\bibitem{wang2012dynamics}
Y.~Wang, G.~Xiao, and J.~Liu, ``Dynamics of competing ideas in complex social
  systems,'' \emph{New Journal of Physics}, vol.~14, no.~1, p. 013015, 2012.

\bibitem{R3}
P.~E. Par{\'e}, J.~Liu, C.~L. Beck, A.~Nedi{\'c}, and T.~Ba{\c{s}}ar,
  ``{Multi-competitive viruses over static and time-varying networks},'' in
  \emph{IEEE American Control Conference}, Seattle, WA, 2017.

\bibitem{R4}
A.~Janson, S.~Gracy, P.~E. Paré, H.~Sandberg, and K.~H. Johansson, ``{Analysis
  of a Networked {SIS} Multi-Virus Model with a Shared Resource},''
  \emph{IFAC-PapersOnLine}, vol.~53, no.~5, pp. 797--802, 2020, 3rd IFAC
  Workshop on Cyber-Physical and Human Systems CPHS 2020.

\bibitem{R5}
A.~Janson, S.~Gracy, P.~E. Par{\'e}, H.~Sandberg, and K.~H. Johansson,
  ``{Networked Multi-Virus Spread with a Shared Resource: Analysis and
  Mitigation Strategies},'' \emph{ArXiv}, vol. abs/2011.07569, 2020.

\bibitem{R6}
P.~E. Par{\'e}, J.~Liu, C.~L. Beck, A.~Nedi{\'c}, and T.~Başar,
  ``{Multi-competitive viruses over time-varying networks with mutations and
  human awareness},'' \emph{Autom.}, vol. 123, p. 109330, 2021.

\bibitem{Bansal2007}
S.~Bansal, B.~Grenfell, and L.~Meyers, ``When individual behaviour mattersl
  homogeneous and network models in epidemiology,'' \emph{Journal Royal
  Society, Interface}, vol.~4, no.~16, pp. 879--891, 2007.

\bibitem{Hochberg1991}
M.~E. Hochberg, ``Non-linear transmission rates and the dynamics of infectious
  disease,'' \emph{Journal of Theoretical Biology}, vol. 153, no.~3, pp.
  301--321, 1991.

\bibitem{Hu2013}
H.~Hu, K.~Nigmatulina, and P.~Eckhoff, ``The scaling of contact rates with
  population density for the infectious disease models,'' \emph{Mathematical
  Biosciences}, vol. 244, no.~2, pp. 125--134, 2013.

\bibitem{Barlow2000}
N.~D. Barlow, ``Non-linear transmission and simple models for bovine
  tuberculosis,'' \emph{Journal of Animal Ecology}, vol.~69, no.~4, pp.
  703--713, 2000.

\bibitem{gan_2013_computer_viruses}
C.~Gan, X.~Yang, W.~Liu, Q.~Zhu, and X.~Zhang, ``An epidemic model of computer
  viruses with vaccination and generalized nonlinear incidence rate,''
  \emph{Applied Mathematics and Computation}, vol. 222, pp. 265--274, 2013.

\bibitem{Liu1987}
W.~Liu, H.~Hetchote, and S.~Levin, ``Dynamical behavior of epidemiological
  models with nonlinear incidence rates.'' \emph{Journal of Mathematical
  Biology}, vol.~25, no.~4, pp. 359--380, 1987.

\bibitem{yang_2015_impact}
L.-X. Yang and X.~Yang, ``The impact of nonlinear infection rate on the spread
  of computer virus,'' \emph{Nonlinear Dynamics}, vol.~82, 05 2015.

\bibitem{yuan_2012_modeling}
H.~Yuan, G.~Liu, and G.~Chen, ``On modeling the crowding and psychological
  effects in network-virus prevalence with nonlinear epidemic model,''
  \emph{Applied Mathematics and Computation}, vol. 219, p. 2387–2397, 11
  2012.

\bibitem{ruan_2013_dynamical}
S.~Ruan and W.~Wang, ``Dynamical behavior of an epidemic model with a nonlinear
  incidence rate,'' \emph{Journal of Differential Equations}, vol. 188, no.~1,
  pp. 135--163, 2003.

\bibitem{HLSmith'17}
H.~L. Smith, ``Monotone dynamical systems: Reflections on new advances and
  applications,'' \emph{Discrete and Continuous Dynamical Systems - A},
  vol.~37, p. 485, 2017.

\bibitem{DeLeenheer2001}
P.~{De Leenheer} and D.~{Aeyels}, ``Stability properties of equilibria of
  classes of cooperative systems,'' \emph{IEEE Transactions on Automatic
  Control}, vol.~46, no.~12, pp. 1996--2001, 2001.

\bibitem{Angeli2003}
D.~{Angeli} and E.~D. {Sontag}, ``Monotone control systems,'' \emph{IEEE
  Transactions on Automatic Control}, vol.~48, no.~10, pp. 1684--1698, 2003.

\bibitem{Bokharaie2010}
V.~S. {Bokharaie}, O.~{Mason}, and M.~{Verwoerd}, ``D-stability and
  delay-independent stability of homogeneous cooperative systems,'' \emph{IEEE
  Transactions on Automatic Control}, vol.~55, no.~12, pp. 2882--2885, 2010.

\bibitem{VanHien2018}
L.~{Van Hien} and H.~{Trinh}, ``Exponential stability of two-dimensional
  homogeneous monotone systems with bounded directional delays,'' \emph{IEEE
  Transactions on Automatic Control}, vol.~63, no.~8, pp. 2694--2700, 2018.

\bibitem{Efimov2013}
D.~{Efimov}, T.~{Raissi}, and A.~{Zolghadri}, ``Control of nonlinear and lpv
  systems: Interval observer-based framework,'' \emph{IEEE Transactions on
  Automatic Control}, vol.~58, no.~3, pp. 773--778, 2013.

\bibitem{Forni2016}
F.~{Forni} and R.~{Sepulchre}, ``Differentially positive systems,'' \emph{IEEE
  Transactions on Automatic Control}, vol.~61, no.~2, pp. 346--359, 2016.

\bibitem{Hirsch-I}
M.~W. Hirsch, ``Systems of differential equations which are competitive or
  cooperative: {I}. limit sets,'' \emph{SIAM Journal on Mathematical Analysis},
  vol.~13, no.~2, pp. 167--179, 1982.

\bibitem{Altafini2013}
C.~{Altafini}, ``Consensus problems on networks with antagonistic
  interactions,'' \emph{IEEE Transactions on Automatic Control}, vol.~58,
  no.~4, pp. 935--946, 2013.

\bibitem{Marco2012}
M.~D. {Marco}, M.~{Forti}, M.~{Grazzini}, and L.~{Pancioni}, ``Limit set
  dichotomy and multistability for a class of cooperative neural networks with
  delays,'' \emph{IEEE Transactions on Neural Networks and Learning Systems},
  vol.~23, no.~9, pp. 1473--1485, 2012.

\bibitem{R2}
M.~Ye, B.~Anderson, and J.~Liu, ``Convergence and equilibria analysis of a
  networked bivirus epidemic model,'' \emph{arXiv preprint arXiv:2111.07507},
  2021.

\bibitem{meyer_textbook}
C.~D. Meyer, \emph{Matrix analysis and applied linear algebra}.\hskip 1em plus
  0.5em minus 0.4em\relax SIAM, 2000, vol.~71.

\bibitem{Berman1994book}
A.~Berman and R.~J. Plemmons, \emph{Nonnegative Matrices in the Mathematical
  Sciences}.\hskip 1em plus 0.5em minus 0.4em\relax SIAM, 1994.

\bibitem{Analysis-Yeh}
J.~Yeh, \emph{Real Analysis}, 2nd~ed.\hskip 1em plus 0.5em minus 0.4em\relax
  WORLD SCIENTIFIC, 2006.

\bibitem{Hirsch-II}
M.~W. Hirsch, ``Systems of differential equations that are competitive or
  cooperative: {II}. convergence almost everywhere,'' \emph{SIAM Journal on
  Mathematical Analysis}, vol.~16, no.~3, pp. 423--439, 1985.

\bibitem{Hirsch-III}
------, ``Systems of differential equations which are competitive or
  cooperative: {III}. competing species,'' \emph{Nonlinearity}, vol.~1, no.~1,
  pp. 51--71, 1988.

\bibitem{Hirsch-IV}
------, ``System of differential equations that are competitive or cooperative:
  {IV}. structural stability in three-dimensional systems,'' \emph{SIAM Journal
  on Mathematical Analysis}, vol.~21, no.~5, p. 1225–1234, 1990.

\bibitem{Hirsch-V}
------, ``Systems of differential equations that are competitive or
  cooperative: {V}. convergence in 3-dimensional systems,'' \emph{Journal of
  Differential Equations}, vol.~80, no.~1, pp. 94 -- 106, 1989.

\bibitem{HLSmith'88}
H.~L. Smith, ``Systems of ordinary differential equations which generate an
  order preserving flow. a survey of results,'' \emph{SIAM Review}, vol.~30,
  no.~1, pp. 87--113, 1988.

\bibitem{HLSmith'90}
H.~L. Smith and H.~R. Thieme, ``Quasi convergence and stability for strongly
  order-preserving semiflows,'' \emph{SIAM Journal on Mathematical Analysis},
  vol.~21, no.~3, pp. 673--692, 1990.

\bibitem{HLSmith'91}
------, ``Convergence for strongly order-preserving semiflows,'' \emph{SIAM
  Journal on Mathematical Analysis}, vol.~22, no.~4, pp. 1081--1101, 1991.

\bibitem{HLSmith'04}
M.~W. Hirsch and H.~L. Smith, ``{Generic Quasi-convergence for Strongly Order
  Preserving Semiflows: A New Approach},'' \emph{Journal of Dynamics and
  Differential Equations}, vol.~16, pp. 433--439, 2004.

\bibitem{MDSbook}
H.~L. Smith, \emph{{Monotone dynamical systems: An introduction to the theory
  of competitive and cooperative systems}}.\hskip 1em plus 0.5em minus
  0.4em\relax American Mathematical Society, 2014.

\bibitem{HS_Co-Op}
\BIBentryALTinterwordspacing
------, ``Is my system of {ODE}s cooperative?'' 2012. [Online]. Available:
  \url{https://math.la.asu.edu/~halsmith/identifyMDS.pdf}
\BIBentrySTDinterwordspacing

\bibitem{Krause_trichotomy}
U.~Krause and P.~Ranft, ``A limit set trichotomy for monotone nonlinear
  dynamical systems,'' \emph{Nonlinear Analysis: Theory, Methods \&
  Applications}, vol.~19, no.~4, pp. 375 -- 392, 1992.

\bibitem{R1}
M.~Ye, J.~Liu, B.~D. Anderson, and M.~Cao, ``{Applications of the Poincare-Hopf
  Theorem: Epidemic Models and Lotka-Volterra Systems},'' \emph{IEEE
  Transactions on Automatic Control}, 2021.

\bibitem{Appendices}
V.~Doshi, S.~Mallick, and D.~Y. Eun, ``Convergence of bi-virus epidemic models
  with non-linear rates on networks - a monotone dynamical systems approach:
  Supplementary material.''

\bibitem{snapnets}
J.~Leskovec and A.~Krevl, ``{SNAP Datasets}: {Stanford} large network dataset
  collection,'' \url{http://snap.stanford.edu/data}, jun 2014.

\bibitem{Perko2001}
L.~Perko, \emph{Differential Equations and Dynamical Systems}, 3rd~ed.\hskip
  1em plus 0.5em minus 0.4em\relax Springer Science \& Business Media, 2001.

\bibitem{ross1996stochastic}
S.~Ross, \emph{Stochastic {P}rocesses}.\hskip 1em plus 0.5em minus 0.4em\relax
  Wiley, 1996.

\end{thebibliography}

\appendices
\section{Basic Definitions and Results from Matrix Theory}\label{matrix theory results}
We first provide some well known results surrounding irreducible square matrices.

\begin{definition}\cite{meyer_textbook}\label{irreducible matrix}
  A square matrix $\mA$ is \textbf{reducible} if there exists a permutation matrix $\mP$ such that $\mP^T\mA\mP$ is a block diagonal matrix. If no such permutation matrix exists, we say that $\mA$ is \textbf{irreducible}.
\end{definition}
\noindent One way to check if a matrix is irreducible is by observing the underlying directed graph, where there is an edge between two nodes only if $a_{ij} \neq 0$. The matrix $A$ is irreducible if and only if this underlying directed graph is strongly connected.
\begin{definition}\cite{Berman1994book}\label{M-matrix}
  A \textbf{M-matrix} is a matrix with non-positive off-diagonal elements with eigenvalues whose real parts are non-negative.
\end{definition}

\noindent We use the following well known result for non-negative, irreducible matrices heavily throughout the paper.

\begin{theorem}(Perron-Frobenius)\label{PF theorem}\cite{meyer_textbook}
  Let $\mA$ be a non-negative, irreducible matrix. Then, $\lambda(\mA)$ is a strictly positive real number, and the corresponding eigenvector $\vecv$ where $\mA\vecv = \lambda(\mA)\vecv$ is also strictly positive. We call $\lambda(\mA)>0$ and $\vecv\gg \0$ the \textbf{PF eigenvalue} and \textbf{PF eigenvector} of the matrix respectively.$\qedsymbol$
\end{theorem}


\noindent The following result is on irreducible M-matrices.

\begin{lemma}\cite{Berman1994book}\label{M matrix lemma}
  Given an irreducible and non-singular M-matrix $\mM$, its inverse $\mM^{-1}$ has strictly positive entries.$\qedsymbol$
\end{lemma}

\section{Definitions and results from ODE literature}\label{ODE results}

We use the following definitions and results from the ODE literature throughout the paper.

\begin{definition}\label{flow definition}
  The `flow' of a dynamical system in a metric space $X$ is a map $\phi: X\!\times\! \R \!\to\! X$ such that for any $x_0 \!\in\! X$ and all $s,t \in \R$, we have $\phi_0(x_0) \!=\! x_0$ and $\phi_s\left( \phi_t(x_0) \right) \!=\! \phi_{t+s}(x_0)$.
\end{definition}

\begin{definition}
  A flow $\phi:X\times \R \to X$ is \textbf{positively invariant} in set $P\subset X$ if for every $x_0 \in P$, $\phi_t(x_0) \in P$ for all $t > 0$.
\end{definition}

\begin{definition}\label{fixed point definiton}
  Given a flow $\phi$, an `equilibrium' or a `fixed point' of the system is a point $x^* \in X$ such that $\{x^*\}$ is a positively invariant set. For the ODE system
  $\dot x = F(x)$,
  we have $F(x^*) = 0$ at the equilibrium.
\end{definition}

\noindent For an equilibrium point $x^*\in X$ we say that the trajectory starting at $x_0\in X$ \textit{converges} to $x^*$ if $\lim_{t \to \infty} \phi_t(x_0) = x^*$. The following result is true for stable fixed points of the ODE system from Definition \ref{flow definition}.

\begin{proposition}\cite{Perko2001} \label{Manifold}
  Let $\mJ F(x_0)$ be the Jacobian of the ODE system evaluated at a fixed point $x_0$ and assume it to be an irreducible matrix. Let $\lambda\left( \mJ F(x_0) \right) < 0$ and suppose the corresponding eigenvalue $\vecv$ is strictly positive ($\vecv \gg \0$). Then, there exists an $\epsilon > 0$ such that $F(x_0 + r\vecv) \ll 0$ for all $r \in (0,\epsilon]$ and $F(x_0 + r\vecv) \gg 0$ for all $r \in (0,-\epsilon]$\footnotemark.$\qedsymbol$
\end{proposition}

\footnotetext{In other words eigenvector $\vecv$ is tangent to the stable manifold of the ODE system at the stable fixed points $x_0$.}

\section{DFR Processes as Non-Linear Recovery Rates} \label{connection to DFR}

In this appendix, we form the connection between \textit{failure rates} from reliability theory \cite{ross1996stochastic}, and the infection duration at any node in SIS type epidemics.
To this end, we start by formally defining the term \textit{failure rate}.

\smallskip
\begin{definition}\label{failure rate} \cite{ross1996stochastic}
Let $T > 0$ be any continuous random variable with distribution $F_T(s) = \prob(T \leq s)$, and density function $f_T(s)$ for all $s>0$, with $\bar F_T(s) = 1-F_T(s) = \prob(T > s)$. Then, the \textit{failure rate} at any given time $s>0$ is defined as
\begin{equation}
    r_T(s) \triangleq \frac{f_T(s)}{\bar F_T(s)}.
\end{equation}
We say $T$ has a decreasing/increasing failure rate (DFR/IFR) if $r_T(s)$ is a decreasing/increasing function of $s>0$.
\end{definition}
\smallskip

When $T$ is the lifetime of a system, the DFR case corresponds to the system aging \textit{negatively}. This means that as time elapses, the residual time (time till the system fails) is more likely to increase rather than decrease. $T$ could also have an interpretation in the context of node recovery. For the linear SIS epidemic model as in \eqref{ode-sis-i}, consider an infected node $i\in\cN$ and define $T \triangleq$ time taken for node $i$ to recover (random), with $f_T(s)$ and $\bar F_T(s)$ as in Definition \ref{failure rate}. Loosely speaking, we can ignore the infection rate terms in \eqref{ode-sis-i} to take a closer look at the recovery process via the ODE
\begin{equation}\label{linear recovery ode}
    \dot x_i(s) = -\delta x_i(s),
\end{equation}
with the initial condition $x_i(0) = 1$ (implying that node $i$ is last infected at time $s=0$). The ODE \eqref{linear recovery ode} has an exact solution for all $s>0$, given by $x_i(s) = e^{-\delta s}.$ This solution allows us to interpret $x_i$ as the cumulative distribution function (CCDF) of an exponential random variable\footnote{When $T \sim \exp(\delta)$, we have $\bar F_T(s) = P(T>s) = e^{-\delta s}$.} with rate $\delta > 0$. Using this interpretation, we have $x_i(s) = P(T>s) = \bar F_T(s)$, and $-\dot x_i(s) = f_T(s)$. \eqref{linear recovery ode} can then be rewritten as
$$r_T(s) = \frac{-\dot x_i(s)}{x_i(s)} = \delta,$$
for any $s>0$. $T$ is thus exponentially distributed, and has a \textit{constant failure rate} (it is both DFR and IFR).

We now consider the case where the random variable $T$ is defined for the more general SIS epidemic model with non-linear recovery rate $q_i(x_i)$ for node $i$.\footnote{Note that this is the special case where $q_i$ is only a function of $x_i$, not of $x_j$ for neighbors $j$ of node $i$.} Ignoring the infection rate terms in \eqref{ode-sis-i-nonlinear} like before, we obtain
\begin{equation}\label{nonlinear recovery ode}
    \dot x_i(s) = -q_i\left(x_i(s)\right),
\end{equation}
retaining the previous interpretation of $x_i$ as the CCDF of $T$. This can be further rearranged to obtain an expression for the failure rate as
$$r_T(s) = \frac{-\dot x_i(s)}{x_i(s)} = \frac{q_i\left(x_i(s)\right)}{x_i(s)}$$
for any $s>0$. From Definition \ref{failure rate} we know $T$ is DFR if $r_T(s)$ is decreasing in $s>0$. Supposing $q_i$ is such that $T$ is indeed DFR, $\log(r_T(s))$ is also decreases in $s$, and we get
$$\frac{d}{ds}\log\left(r_T(s)\right) = \frac{q_i'(x_i(s)) \dot x_i(s) }{q(x_i(s))} - \frac{\dot x_i(s)}{x_i(s)} \leq 0,$$
where $q_i'(x_i(s))$ denotes the derivative with respect to $x_i$. Since $\dot x_i(s) = -q_i(x_i(s))$ from \eqref{nonlinear recovery ode} and $q_i'(x(s)) \geq 0$ from (A3), rearranging the previous equation gives us following the condition for $T$ to be DFR
\begin{equation}\label{DFR condition}
    x_i q_i'(x_i) - q_i(x_i) \geq 0.  
\end{equation}
In \eqref{DFR condition}, the $(s)$ notation has been suppressed for clarity. Since $q_i(0) = 0$, the convexity of $q_i$ with respect to $x_i$ implies \eqref{DFR condition}. 

Roughly speaking, the DFR case (which also includes linear recovery rates as in \eqref{ode-sis-i}) is a subclass of recovery rate functions $q_i(\vx)$ satisfying assumptions (A1)--(A5). Even though the above steps may not be exact, they provide intuition on how infections which fester and grow worse with time form part of our modelling assumptions in Section \ref{nonlinear epidemic models}.

\section{Results from MDS and Cooperative Systems}\label{MDS}

\begin{definition}\cite{Hirsch-I, HLSmith'88,HLSmith'17}\label{monotone and co-operative}
  A flow $\phi$ is said to be \textbf{monotone} if for all $\vx, \vy \in \R^n$ such that $\vx \leq_K \vy$ and any $t\geq0$, we have
  $\phi_t(\vx) \leq_K \phi_t(\vy).$

  If the flow represents the solution of an ODE system, we say that the ODE system is \textbf{co-operative}.
\end{definition}

\begin{definition}\label{irreducible ode}
  Consider the system \eqref{ode sys} and let $\mJ F(\vx)\! \triangleq\! \left[{df_i(\vx)}/{dx_j}\right]$ be the Jacobian of the right hand side evaluated at any point $\vx \! \in \! \R^n$. We say that \eqref{ode sys} is an \textbf{irreducible ODE} in set $D \in \R^n$ if for all $\vx \in D$, $\mJ F(\vx)$ is an irreducible matrix.
\end{definition}
\begin{definition}\cite{HLSmith'88,MDSbook,HLSmith'17}\label{strongly monotone}
  The flow $\phi$ is said to be \textbf{strongly monotone} if it is monotone, and for all $\vx,\vy \in \R^n$ such that $\vx <_K \vy$, and time $t \geq 0$, we have
  $\phi_t(\vx) \ll_k \phi_t(\vy).$
\end{definition}
\begin{theorem}\label{SM ODE}\cite{HLSmith'88,MDSbook,HLSmith'17}
  Let \eqref{ode sys} be irreducible and co-operative in some set $D \subset \R^n$. Then the solution $\phi$ (restricted to $t \geq 0$) is strongly monotone.$\qedsymbol$
\end{theorem}
As part of the main result of monotone dynamical systems, trajectories of strongly monotone systems, starting from almost anywhere (in the measure theoretic sense) in the state space, converge to the set of equilibrium points \cite{Hirsch-II,HLSmith'91,HLSmith'04,HLSmith'17}. However, often the systems are strongly monotone only in the interior of the state spaces instead of the entirety of the state space. In such cases, the following results are useful.

\begin{proposition}(Proposition 3.2.1 in \cite{MDSbook})\label{invariance prop}
Consider the ODE system \eqref{ode sys} which is cooperative in a compact set $D\subset \R^n$ with respect to some cone-ordering, and let $<_r$ stand for any of the order relations $\leq_K, <_K, \ll_K$. Then, $P_+ \!\triangleq\! \left\lbrace \vx \!\in\! D ~|~ \0 \!<_r\! F(\vx) \right\rbrace$ and $P_- \!\triangleq\! \left\lbrace \vx \!\in\! D ~|~ F(\vx) \!<_r\! \0 \right\rbrace$ are positively invariant, and the trajectory $\left\lbrace \phi_t(\vx) \right\rbrace_{t \geq 0}$ for any point $\vx \!\in\! P_+$ or $\vx \!\in\! P_-$ converges to an equilibrium.$\qedsymbol$
\end{proposition}

\begin{theorem}(Theorem 4.3.3 in \cite{MDSbook})\label{existence of another fp}
Let \eqref{ode sys} be cooperative (with respect to some cone-ordering $\leq_K$) in a compact set $D \subset \R^n$ and let $\vx_0 \in D$ be an equilibrium point. Suppose that $s \triangleq \lambda(\mJ F(\vx_0))>0$ (i.e. $\vx_0$ is an unstable fixed point) and there is an eigenvector $\vecv \gg_K \0$ such that $\mJ F(\vx_0) \vecv = s \vecv$. Then, there exists $\epsilon_0 \in (0,\epsilon]$ and another equilibrium point $\vx_e$ such that for each $r \in (0,\epsilon_0]$, the solution $\phi_t(\vx_r)$ has the following properties:
\begin{itemize}
\item[(1)]
$\vx_r \!\ll_K\! \phi_{t_1}(\vx_r)\! \ll_K\! \phi_{t_2}(\vx_r)\! \ll_K \!\vx_e$, for any $0\!<\!t_1\!<\!t_2$.
\item[(2)]
${d \phi_t(\vx_r)}/{dt} \gg_K \0$, for any $t>0$.
\item[(3)]
$\phi_t(\vx_r) \rightarrow \vx_e$, as $t \rightarrow \infty$.$\qedsymbol$
\end{itemize}
\end{theorem}

\section{Proofs of the results in Section \ref{monotonicity of epidemic models}}\label{single virus proofs}

\begin{IEEEproof}[\textbf{Proof of Proposition \ref{prop: single sis cooperative}}]
  To prove that system \eqref{ode-sis-nonlinear} is co-operative with respect to the positive orthant, we show that it satisfies Kamke's condition in \eqref{Kamke for positive}. Differentiating the right hand side of \eqref{ode-sis-i-nonlinear} with respect to $x_j$, we get
  \begin{align*}
    \frac{\partial \bar f_i (\vx)}{\partial x_j} = (1-x_i)\frac{\partial f_i(x)}{\partial x_j} = \frac{\partial q_i(\vx)}{\partial x_j}.
  \end{align*}
  This corresponds to the $(ij)$'th off-diagonal entry of the Jacobian $\mJ_{\bar F}(\vx)$ evaluated at $\vx \in [0,1]^N$. It is non-negative for any $i\neq j \in \cN$ since $(1-x_i)\geq0$ and due to assumption (A3), and the ODE \eqref{ode-sis-nonlinear} is therefore co-operative in $[0,1]^N$ with respect to the regular cone ordering.

  From assumption (A3), $\mJ_{\bar F}(\vx)_{ij}$ is also strictly positive for any $\vx \in (0,1)^N$ whenever $a_{ij}>0$. This means that $\mJ_{\bar F}(\vx)$, and as a consequence the ODE system, is irreducible for any $\vx \in (0,1)^N$.
\end{IEEEproof}\smallskip

To derive the convergence properties of the non-linear $SIS$ model, we make use of a result form \cite{Krause_trichotomy}, rewritten below in a simpler form suitable for our setting.
\smallskip
\begin{theorem}\label{limit set trichotomy}(Theorem 4 in \cite{Krause_trichotomy})
Consider a generic ODE system \eqref{ode sys} invariant to some subset $S\subset \R^N_+$, and let $\mJ_{\bar F}$ stand for its Jacobian matrix. Suppose that:
\begin{itemize}
  \item[(C1)] $f_i(\vx) \geq 0$ for all $\vx \geq 0$ with $x_i = 0$;
  \item[(C2)] for all $\vx \gg \0$ in $S$, $\alpha \in (0,1)$, it satisfies $\mJ_{\bar F}(\vx)_{ij} \leq \mJ_{\bar F}(\alpha \vx)_{ij}$ for all $i,j\in\cN$, with strict inequality for at least one pair of $i,j$;
  \item[(C3)] for all $\vu \ll \vw$ in $S$, it satisfies $\mJ_{\bar F}(\vw) \leq \mJ_{\bar F}(\vu)$;
  \item[(C4)] it is co-operative in $S$ with respect to the regular ordering relation, and irreducible in $\text{Int}(S)$.
\end{itemize}
Then, exactly one of the following outcomes occurs:
\begin{enumerate}[(i)]
  \item $\phi_t(\vx)$ is unbounded for all $\vx \in S \setminus \{ \0 \}$;
  \item $\phi_t(\vx) \rightarrow \0$ as $t\rightarrow\infty$, for all $\vx \in S \setminus \{ \0 \}$;
  \item There exists a unique, strictly positive fixed point $\vx^*\gg\0$ such that $\phi_t(\vx) \rightarrow \vx^*$ as $t\rightarrow\infty$, for all $\vx \in S \setminus \{ \0 \}$.$\qedsymbol$
\end{enumerate}
\end{theorem}\smallskip

We now use the above to prove Theorem \ref{nonlinear sis-conditions}.

\smallskip
\begin{IEEEproof}[\textbf{Proof of Theorem \ref{nonlinear sis-conditions}}]
  We prove Theorem \ref{nonlinear sis-conditions} by showing that it satisfies conditions (C1)-(C4) of Theorem \ref{limit set trichotomy}, and then performing stability analysis to evaluate conditions for each of the three possible outcomes therein.

  From Proposition \eqref{prop: single sis cooperative}, we know that \eqref{ode-sis-nonlinear} already satisfies (C4). The right hand side of \eqref{ode-sis-i-nonlinear} satisfies (C1) because $q_i(x_i) = 0$ when $x_i=0$, and because $(1-x_i)$ and $f_i(\vx)$ are all non-negative for any $\vx \in [0,1]^N$. To check whether (C2) and (C3) is satisfied, observe that from assumptions (A2)--(A5), we have
  \begin{align}
    \mJ_F(\vu) > \mJ_F(\vw) \label{assumption consequence F}\\
    \mJ_Q(\vu)< \mJ_Q(\vw) \label{assumption consequence Q}
  \end{align}
  for all $\vu < \vw$.\footnote{Here, the ordering between matrices $\mM^a < \mM^b$ means $\mM^a_{ij} \leq \mM^b_{ij}$ with the inequality being strict for at least one pair of $i,j$.} Here, $\mJ_Q$ is a diagonal matrix since $\partial q_i / \partial x_j = 0$ for all $i \neq j \in \cN$.

  Denote by $\mJ_{\bar F}$ the Jacobian matrix of system \eqref{ode-sis-nonlinear}. Note that for any point $\vx \in [0,1]^N$, we have
  \begin{equation}\label{jacobian-single}
      \mJ_{\bar F}(\vx) = \text{diag}(\ones - \vx)\mJ_{F}(\vx) - \text{diag}\left(F(\vx)\right) - \mJ_Q(\vx)
  \end{equation}
  Combining the above with \eqref{assumption consequence F} and \eqref{assumption consequence Q}, we have for any points $\vu < \vw$ that
  \begin{align*}
    \mJ_{\bar F}(\vu)   &= \text{diag}(\ones - \vu)\mJ_{F}(\vu) - \text{diag}\left(F(\vu)\right) - \mJ_Q(\vu) \\
                        &> \text{diag}(\ones - \vw)\mJ_{F}(\vw) - \text{diag}\left(F(\vu)\right) - \mJ_Q(\vw) \\
                        &\geq \text{diag}(\ones - \vw)\mJ_{F}(\vw) - \text{diag}\left(F(\vw)\right) - \mJ_Q(\vw) \\
                        &= \mJ_{\bar F}(\vw),
  \end{align*}
  where the first inequality is due to $(\ones - \vu) > (\ones - \vw)$ and \eqref{assumption consequence F} and \eqref{assumption consequence Q}. The second inequality is from the non-negativity and monotonicity assumptions (A2) and (A3) implying $F(\vu)\leq F(\vw)$. Since $\mJ_{\bar F}(\vu) > \mJ_{\bar F}(\vw)$ for any $\vu < \vw$, this is enough to satisfy both conditions (C2) and (C3).

  Since system  \eqref{ode-sis-nonlinear} satisfies (C1)--(C4), Theorem \ref{limit set trichotomy} applies. Since the system is invariant in $[0,1]^N$, which is a bounded subset of $\R^N$, outcome (i) of Theorem \ref{limit set trichotomy} never occurs. From assumption (A1), the vector $\0 = [0,\cdots,0]^T$ (the virus-free equilibrium) is always a fixed point of the system. We now find conditions under which trajectories of \eqref{ode-sis-nonlinear} starting from anywhere in $[0,1]^N \setminus \{ \0 \}$ converge to either zero, or to a unique strictly positive fixed point (outcomes (ii) and (iii) in Theorem \ref{limit set trichotomy} respectively), by check the stability properties of the system.

  The virus-free fixed point zero is unstable \cite{Perko2001} when $\lambda(\mJ_{\bar F}(\0)) = \lambda(\mJ_F(\0) - \mJ_Q(\0)) \leq 0$. Under this condition, outcome (ii) in Theorem \ref{limit set trichotomy} is not possible, and there exists a unique, strictly positive fixed point $\vx^* \gg \0$ which is globally asymptotically stable in $[0,1]^N \setminus \{ \0 \}$. Conversely when zero is a stable fixed point, that is when $\lambda(\mJ_{\bar F}(\0)) = \lambda(\mJ_F(\0) - \mJ_Q(\0)) > 0$, it is globally attractive.
\end{IEEEproof}\smallskip

\section{Proofs of the Main Results}\label{proof of the results}

Throughout this Section, we use $\phi_t(\vx_0,\vy_0)$ to represent the solution of \eqref{eq:biSIS-nonlinear} at time $t \geq 0$, starting from $(\vx_0,\vy_0) \in D$. We will need the following results to prove the theorems from Section \ref{convergence and coexistence}.

\begin{proposition}\label{convergence to Z}
  Starting from any point $D \setminus \left\{ (\0,\0) \right\}$, trajectories of \eqref{eq:biSIS-nonlinear} converge to the set
  \begin{equation*}
      Z \triangleq \left\{ (\vu,\vw) \in D ~|~ (\0,\vy^*)\leq_K (\vu,\vw) \leq_K (\vx^*, \0) \right\}.
  \end{equation*}
\end{proposition}
\begin{IEEEproof}
  For any $(\vr,\vs) \in D\setminus \{(\0,\0)\}$, there exists points $\vx,\vy \in [0,1]^N$ such that $(\0,\vy) \leq_K (\vr,\vs) \leq_K (\vx,\0)$. Then, from Definition \ref{monotone and co-operative} of a monotone system, we have $\phi_t(\0,\vy) \leq_K \phi_t(\vr,\vs) \leq_K \phi_t(\vx,\0)$ for any $t > 0$. Since $\phi_t(\vx,\0)\rightarrow(\vx^*,\0)$ and $\phi_t(\0,\vy) \rightarrow (\0,\vy^*)$, we get $(\0,\vy^*) \leq_K \lim_{t \rightarrow \infty} \phi_t(\vr,\vs) \leq_K  (\vx^*,\0)$. Thus the trajectory $\left\{\phi_t(\vr,\vs)\right\}_{t \geq 0}$ converges to $Z$, completing the proof.
\end{IEEEproof}\smallskip

Since the set $Z$ depends on $\vx^*$ and $\vy^*$, the fixed points of systems \eqref{sis-x} and \eqref{sis-y}, and we can determine when these fixed points are positive or zero, Proposition \ref{convergence to Z} helps us to quickly point out a subset of the state space to which trajectories starting from any point in $D\!\setminus\! \left\{ (\0,\0) \right\}$ converge.

\smallskip
\begin{IEEEproof}[\textbf{Proof of Theorem \ref{theorem virus free}}]
  When $\lambda \left( \mJ_G(\0) - \mJ_R(\0) \right) \!\leq\! 0$ and $\lambda \left( \mJ_H(\0) - \mJ_S(\0) \right) \!\leq\! 0$, we know from Theorem \ref{nonlinear sis-conditions} that $\vx^* = \vy^* = 0$. Therefore, trajectories of \eqref{eq:biSIS-nonlinear} starting from any point in $D\setminus \left\{ (\0,\0) \right\}$ converge to the set $Z \triangleq \left\{ (\vu,\vw) \in D ~|~ (\0,\0)\leq_K (\vu,\vw) \leq_K (\0, \0) \right\} = \left\{ (\0,\0) \right\}$. Hence, the virus-free equilibrium is globally asymptotically stable in $D$, which completes the proof.
\end{IEEEproof}\smallskip

Proposition \ref{convergence to Z} can also be applied to show that $(\vx^*,\0)$ where $\vx^* \!\gg\! \0$ is globally attractive when $\lambda \left( \mJ_G(\0) \!-\! \mJ_R(\0) \right) \!>\! 0$ and $\lambda \left( \mJ_H(\0) \!-\! \mJ_S(\0) \right) \!\leq\! 0$. This is because from Theorem \ref{nonlinear sis-conditions}, we know that $\vx^* \!\gg\! \0$ and $\vy^* \!=\! \0$. We then have $Z \triangleq \left\{ (\vu,\vw) \in D ~|~ (\0,\0)\leq_K (\vu,\vw) \leq_K (\vx^*, \0) \right\}$, implying that the system \eqref{eq:biSIS-nonlinear} ultimately reduces to the single $SIS$ system \eqref{sis-x}, which we know globally converges to $\vx^*$. By a symmetric argument, we also have that $(\0,\vy^*)$ where $\vy^* \!\gg\! \0$ is globally attractive when $\lambda \left( \mJ_G(\0) \!-\! \mJ_R(\0) \right) \!\leq\! 0$ and $\lambda \left( \mJ_H(\0) \!-\! \mJ_S(\0) \right) \!>\! 0$. Therefore these cases are easily analyzed by applying Proposition \ref{convergence to Z} in conjunction with Theorem \ref{nonlinear sis-conditions}. In terms of the linear bi-virus model whose parameters are easier to visualize, values of $\tau_1$ and $\tau_2$ which satisfy these conditions, lie in regions R2 and R3 of Figure \ref{all regions}(b) and we henceforth exclude them from our analysis, considering only those values of $\tau_1$ and $\tau_2$ for which $\tau_1\lambda(\mA)\!>\!1$ and $\tau_2\lambda(\mB)\!>\!1$ always holds; equivalently considering only the cases where $\lambda \left( \mJ_G(\0) - \mJ_R(\0) \right) > 0$ and $\lambda \left( \mJ_H(\0) - \mJ_S(\0) \right) > 0$ always hold for nonlinear infection and recovery rates. Thus, $\vx^*$ and $\vy^*$ are henceforth implied to be strictly positive vectors.

Before formally proving Theorems \ref{theorem wta} and \ref{theorem coexistence}, we provide some additional constructions and notations which will help simplify the proofs. As in the proof of Theorem \ref{nonlinear sis-conditions}, the Jacobians $\mJ F^x(\vx)$ and $\mJ F^y(\vy)$ of systems \eqref{sis-x} and \eqref{sis-y}, respectively, are
\begin{align*}
  \mJ F^x(\vx) &= \text{diag}(\ones \!-\! \vx)\mJ_G(\vx) - \text{diag}(G(\vx)) - \mJ_R(\vx), \\
  \mJ F^y(\vy) &= \text{diag}(\ones \!-\! \vy)\mJ_H(\vy) - \text{diag}(H(\vy)) - \mJ_S(\vy),
\end{align*}
\noindent for all $\vx,\vy \in [0,1]^N$. Now recall the Jacobian $\mJ_{\bar G \bar H}(\vx,\vy)$ of the bi-virus ODE \eqref{eq:biSIS-nonlinear} from \eqref{Jacobian bi-virus}. When evaluated at $(\vx^*,\0)$ and at $(\0,\vy^*)$, we get
\begin{equation}\label{Jacobian wta x}
  \begin{split}
    \mJ_{\bar G \bar H}(\vx^*,\0)=
    \begin{bmatrix}
    \mJ F^x(\vx^*)      & \mK \\
         \0           & \mJ_y
    \end{bmatrix}
  \end{split}
\end{equation}
\noindent where $\mK \!=\! -\text{diag}(G(vx^*))$, $\mJ_y \!=\! \text{diag}(\ones \!-\! \vx^*)\mJ_H(\0) \!-\! \mJ_S(\0)$, and
\begin{equation}\label{Jacobian wta y}
  \begin{split}
    \mJ_{\bar G \bar H}(\0,\vy^*)=
    \begin{bmatrix}
    \mJ_x          & \0 \\
     \mL       & \mJ F^y(\vy^*)
    \end{bmatrix}
  \end{split}
\end{equation}
\noindent where $\mL \!=\! -\text{diag}(H(\vy^*))$, $\mJ_x \!=\! \text{diag}(\ones \!-\! \vy^*)\mJ_G(\0) \!-\! \mJ_R(\0)$. This leads us to the following proposition, where the ordering $\leq_K$ ($<_K,\ll_K$) stands for the south east cone-ordering.

\smallskip
\begin{proposition}\label{positive ev}
  When $\lambda \!\left( \mS_{\vy^*}\mJ_G(\0) \!-\! \mJ_R(\0) \right) \!>\! 0$, we have $\lambda\left( \mJ_{\bar G \bar H}(0,\vy^*) \right) \!=\! \lambda(\mJ_x) \!>\! 0$, and the corresponding eigenvector $(\vu,\vecv) \!\in\! \R^{2N}$ of $\mJ_{\bar G \bar H}(0,\vy^*)$ satisfies $(\vu,\vecv) \!\gg_K\! (\0,\0)$.
\end{proposition}\smallskip
\begin{IEEEproof}
  First, recall that $\vy^* \gg \0$ is the asymptotically stable fixed point of \eqref{sis-y}. This implies that the real parts of all eigenvalues of the Jacobian  $\mJ F^y(\vy^*)$ of \eqref{sis-y} evaluated at $\vy^*$ are negative. Since $\mJ F^y(\vy^*)$ is an irreducible matrix as discussed in Section \ref{monotonicity bi-virus}, with non-negative off-diagonal elements, its PF eigenvalue (obtained by perturbing with a large multiple of the identity matrix) is real and negative, that is $\lambda\left(\mJ F^y (\vy^*)\right)<0$.

  From the assumption, we have $\lambda \left( \mS_{\vy^*}\mJ_G(\0) - \mJ_R(\0) \right) = \lambda(\mJ_x) > 0$. Since $\mJ_{\bar G \bar H}(\0,\vy^*)$ is a block triangle matrix, we have $\lambda\left(\mJ_{\bar G \bar H}(\0,\vy^*)\right) \!=\! \max\! \left\{ \lambda(J_x), \lambda\left(\mJ F^y (\vy^*)\right) \right\}$, and since $\lambda\left(\mJ F^y (\vy^*)\right) < 0$, we obtain $\lambda\left(\mJ_{\bar G \bar H}(\0,\vy^*)\right) = \lambda(\mJ_x) > 0$. Then, the corresponding eigenvector $(\vu,\vecv)$ satisfies
  \begin{equation*}
    \mJ_x \vu \!=\! \lambda(\mJ_x)\vu ~~~~\text{and}~~~~
    \mL\vu \!+\! \mJ F^y(\vy^*)\vecv \!=\! \lambda(\mJ_x)\vecv.
  \end{equation*}
  \noindent From the first equation, we can tell that $\vu$ is the eigenvector of $\mJ_x$ corresponding to its PF eigenvalue, and thus satisfies $\vu \!\gg\! \0$. Now recall that $\mJ F^y(\vy^*)$ had eigenvalues with strictly negative real parts. $ \lambda(\mJ_x)\eye \!-\!\mJ F^y(\vy^*)$ is then a matrix with eigenvalues having strictly positive real parts (since $\lambda(\mJ_x)\!>\!0$). The matrix $\mM \triangleq \lambda(\mJ_x) \eye \!-\! \mJ F^y(\vy^*)$ is then, by Definition \ref{M-matrix}, an M-matrix. By construction, it is also irreducible and invertible and from Lemma \ref{M matrix lemma}, we obtain that $\mM^{-1}$ is a (strictly) positive matrix. The second equation in the above can then be rewritten as $\vecv = \mM^{-1}\mL\vu \ll \0$, where the inequality is because $\mL \!=\! -\text{diag}(H(\vy^*))$ has strictly negative diagonal elements ($H(\vy^*)$ being positive from assumptions (A2) and (A3)). Therefore, since $\vu \gg \0$ and $\vecv \ll \0$, we have $(\vu,\vecv) \gg_K \0$, completing the proof.
\end{IEEEproof}\smallskip

The intention behind introducing Proposition \ref{positive ev} was to satisfy the assumptions of Theorem \ref{existence of another fp}. In particular, when $\lambda \left( \mS_{\vy^*}\mJ_G(\0) - \mJ_R(\0) \right) \!>\! 0$, $(0,\vy^*)$ is an unstable fixed point; by Proposition \ref{positive ev} and Theorem \ref{existence of another fp}, there exists an $\epsilon_1 > 0$ and another fixed point $(\vx_e,\vy_e)$ such that for any point $(\vx_r,\vy_r) \triangleq (\0,\vy^*) + r(\vu,\vecv)$ where $r \in (0,\epsilon_1]$, we have
\begin{equation*}\label{monotone seq to xe,ye}
  \begin{split}
    (0,\vy^*) \!\ll\! (\vx_r,\vy_r) \!\ll_K\! \phi_t(\vx_r,\vy_r)
                                \!\ll_K\! \phi_s(\vx_r,\vy_r) \!\leq_K\! (\vx^*,\0)
  \end{split}
\end{equation*}
\noindent for all $s \!>\! t \!>\! 0$. Moreover, for all $(\vx,\vy)$ such that $(\0,\vy^*) \!\ll_K\! (\vx,\vy) \!\leq_K\! (\vx_e,\vy_e)$, there exists an $r\!\in\!(0,\epsilon]$ sufficiently small such that $(\vx_r,\vy_r) \!\leq_K\! (\vx,\vy) \!\leq_K\! (\vx_e,\vy_e)$. Since $\phi_t(\vx_r,\vy_r) \!\rightarrow\! (\vx_e,\vy_e)$, monotonicity implies $\phi_t(\vx,\vy) \!\rightarrow\! (\vx_e,\vy_e)$ as $t\!\to\!\infty$.

Now, we can either have $(\vx_e,\vy_e) \!=\! (\vx^*,\0)$, which occurs when $(\vx^*,\0)$ is the other stable fixed point of \eqref{eq:biSIS-nonlinear}, or $(\vx_e,\vy_e) \!=\! (\hat \vx, \hat \vy) \!\gg\! \0$ which occurs when $(\vx^*,\0)$ is an unstable fixed point. Note that  $(\vx^*,\0)$ is stable (unstable) if and only if $\lambda \left( \mS_{\vy^*}\mJ_G(\0) - \mJ_R(\0) \right) \!\leq\! 0$ ($ >\! 0$). We will talk about both these possibilities one by one and exploring these will eventually lead to Theorems \ref{theorem wta} and \ref{theorem coexistence}. But before we do that, we first prove the following proposition about convergence to the fixed point $(\vx_e,\vy_e)$ (whichever of the two it may be).

\smallskip
\begin{proposition}\label{convergence to x_e,y_e}
  Trajectories of the system \eqref{eq:biSIS} starting from any point $(\vx,\vy)$ such that $(\0,\vy^*) <_K (\vx,\vy) \leq_K (\vx_e,\vy_e)$ converge to $(\vx_e,\vy_e)$.$\qedsymbol$
\end{proposition} \smallskip
\begin{IEEEproof}
  Recall that we already know that for all $(\0,\vy^*) \!\ll_K\! (\vx,\vy) \!\leq\! (\vx^*,\vy^*)$, $\phi_t(\vx,\vy) \!\rightarrow\! (\vx^*,\0)$. We would however like to show this for all $(\vx,\vy)\in Z\setminus(\0,\vy^*)$, that is even when $(\vx,\vy)$ satisfies $(\0,\vy^*) <_K (\vx,\vy) \leq (\vx^*,\0)$. To do this, we create a set of points which converge to $(\vx^*,\0)$, just like we created $(\vx_r,\vy_r)$ before, and then use a monotonicity argument to show convergence to $(\0,\vy^*)$ of trajectories starting for all points $(\vx,\vy)$ satisfying $(\0,\vy^*) <_K (\vx,\vy) \leq (\vx^*,\0)$.

  Recall that, $\vy^*$ is an asymptotically stable fixed point of \eqref{sis-y}, and from the proof of Proposition \ref{positive ev} we know that $\lambda\left(\mJ F^y (\vy^*)\right)<0$. Let $\vw \gg \0$ be the corresponding PF eigenvector. Then by Proposition \ref{Manifold}, there exists an $\epsilon_2 > 0$ such that for all $s \in (0,\epsilon_2]$, $F^y(\vy^* + s \vw) \ll \0$. We can then define points $(\vx_r,\vy_s) \triangleq (r\vu,\vy^* + s\vw)$ for any $r \in (0,\epsilon_1]$ and $s \in (0,\epsilon_2]$, where $\vu\gg \0$ is the eigenvector of $\mJ_x$ from Proposition \ref{positive ev}. We will first show that trajectories starting from these points converge to $(\vx_e,\vy_e)$. By rearranging the terms of \eqref{eq:biSIS-nonlinear}, we can rewrite it as
  \begin{align*}
    \dot \vx =& ~\text{diag}(\ones - \vy^*)G(\vx) - R(\vx) + \text{diag}(\vy^*-\vx-\vy)G(\vx) \\
             =& ~\text{diag}(\ones - \vy^*)\mJ_G(\0)\vx - \mJ_R(\0)\vx \\
             &+ \text{diag}(\vy^*-\vx-\vy)G(\vx) + O\left(\|\vx\|^2\right) \\
             =& ~\mJ_x\vx + O\left(\| \vx \|\left[\| \vy - \vy^* \| + \| \vx \|\right] \right), \\
    \dot \vy =& ~\text{diag}(\ones - \vy)H(\vy) - S(\vy) - \text{diag}(\vx)H(\vy) \\
             =& ~F^y(\vy) + O\left(\| \vy \|\right),
  \end{align*}
  \noindent for all $(\vx,\vy) \in D,$\footnotemark where the first equality is from a Taylor series expansion of $G$ and $R$ around $\0$. For any point $(\vx_r,\vy_s) = (r\vu,\vy^* + s\vw)$, the above equations can be written as
  \footnotetext{Here, $O(x)$ is used to represent terms which satisfy $O(x) \to 0$ as $x \to 0$.}
  \begin{align*}
    \dot \vx &= r \lambda(\mJ_x) \vu + rO\left(\| \vu \|\left[s\| \vw \| + r\| \vu \|\right] \right) \\
             &= r \left[\lambda(\mJ_x)\vu + O(r+s) \right]\ \\
    \dot \vy &= F^y(\vy^* + s\vw) + O\left(\| s \vy \|\right) \\
             &= F^y(\vy^* + s\vw) + O\left( s \right).
  \end{align*}
  \noindent For sufficiently small $r$ and $s$, we have $\dot \vx \gg \0$ (since $\lambda(\mJ_x) >0$ and $\vu \gg \0$) and $\dot \vy \ll \0$ (since $F^y(\vy^* + s \vw) \ll \0$ for all $s\in(0,\epsilon_2]$). This satisfies the conditions for Proposition \ref{invariance prop}, and trajectories starting from such points will be monotonically increasing (according to the south-east cone ordering), eventually converging to the fixed point $(\vx_e,\vy_e)$.

  Now see that for any point $(\vx,\vy)$ such that $(\0,\vy^*) \!<_K\! (\vx,\vy) \!\leq_K\! (\vx_e,\vy_e)$, where $\vx \!>\! \0$ and $\vy \!\leq\! \vy^*$, by the nature of the ODE system \eqref{eq:biSIS} all zero entries of the $\vx$ term will eventually become positive (if it isn't already). Therefore, there exists a time $t_1>0$ such that $\vx(t_1) \!\gg\! \0$, and there exist $r,s$ small enough such that $(\vx_r,\vy_s) \!\ll_K\! \phi_{t_1}(\vx,\vy) \!\leq_K\! (\vx_e,\vy_e)$. Again by monotonicity, since $\phi_t(\vx_r,\vy_s) \rightarrow (\vx_e,\vy_e)$, we have $\phi_{t+t_1}(\vx,\vy)\rightarrow(\vx_e,\vy_e)$ as $t \rightarrow \infty$, completing the proof.
\end{IEEEproof}\smallskip

We now consider the case where $(\vx_e,\vy_e) \!=\! (\vx^*, \0)$ and give the proof for Theorem \ref{theorem wta}. We prove it only for when $\tau_1\lambda(\mS_{\vy^*} \mA ) \!>\! 1$ and $\tau_2\lambda(\mS_{\vx^*} \mB) \!\leq\! 1$, since the other case follows by a symmetric argument.

\smallskip
\begin{IEEEproof}[\textbf{Proof of Theorem \ref{theorem wta}}]
  When $\lambda \left( \mJ_H(\0) \!-\! \mJ_S(\0) \right) \!\leq\! 0$, $(\vx^*,\0)$ is a stable fixed point of system \eqref{eq:biSIS-nonlinear}, since all eigenvalues of $\mJ_{\bar G \bar H}(\vx^*,\0)$ have non-positive real parts, and we have $(\vx_e,\vy_e) = (\vx^*,\0)$. Proposition \ref{convergence to x_e,y_e} then implies that trajectories starting from all points in $Z\setminus \left\{ (\0,\vy^*) \right\}$ converge to $(\vx^*,\0)$. According to Proposition \ref{convergence to Z}, trajectories starting from all points $(\vx,\vy)\in B_x$ in the system eventually enter the set $Z$, thereby eventually converging to $(\vx^*,\0)$, giving us global convergence in $B_x$.
\end{IEEEproof}\smallskip

Similarly, we use Propositon \ref{convergence to x_e,y_e} to prove Theorem \ref{theorem coexistence}.

\smallskip
\begin{IEEEproof}[\textbf{Proof of Theorem \ref{theorem coexistence}}]
  When $\lambda \left( \mJ_G(\0) \!-\! \mJ_R(\0) \right) \!>\! 0$ and $\lambda \left( \mJ_H(\0) \!-\! \mJ_S(\0) \right) \!>\! 0$, both $(\0,\vy^*)$ and $(\vx^*,\0)$ are unstable fixed points, and $(\vx_e,\vy_e)$ takes the form of a positive fixed point $(\hat \vx, \hat \vy) \gg \0$ (it cannot be $(\vx^*,\0)$, which is unstable). Then from Proposition \ref{convergence to x_e,y_e}, it attracts trajectories beginning from all points $(\vx,\vy)$ satisfying $(\0,\vy^*) \!<_K\! (\vx,\vy) \!\leq_K\! (\hat \vx,\hat \vy)$.

  Similarly, we have a symmetric result beginning from $\tau_2\lambda(\mS_{\vx^*} \mB) \!>\! 1$ (symmetric to Proposition \ref{positive ev} which assumes $\tau_1\lambda(\mS_{\vy^*} \mA ) \!>\! 1$ instead), and we can say that there exists another fixed point $(\bar \vx, \bar \vy) \!\gg\! \0$ which attracts all points $(\vx,\vy)$ satisfying $(\bar \vx, \bar \vy) \!\leq_K\! (\vx,\vy)\!<_K\!(\vx^*,\0)$. By construction, we then have $(\hat \vx , \hat \vy) \!\leq_K\! (\bar \vx, \bar \vy)$, with the possibility of being equal.

  To prove global convergence of the system to the set $S \!=\! \left\{ (\vx_e,\vy_e) \!\in\! E ~|~ (\hat \vx,\hat \vx) \!\leq_K\! (\vx_e,\vy_e) \!\leq_K\! (\bar \vx,\bar \vy)\right\}$, observe first that as part of the proof of Proposition \ref{convergence to x_e,y_e} we showed that for trajectories starting from any point $(\vx,\vy)$ in the state space, there exists $r\!>\!0$ and $s\!>\!0$ small enough, and $t_1\!>\!0$ such that $(\vx_r,\vy_s) \!\ll_K\! \phi_{t_1}(\vx,\vy) \!\leq_K\! (\hat \vx,\hat \vy)$ where $(\vx_r,\vy_s)$ is a point very close to $(\vx^*,\0)$. By a parallel argument, we can find a similar point $(\vx_p,\vy_q)$ very close to $(\0,\vy^*)$ and a time $t_2$ such that $(\bar \vx,\bar \vy) \!\leq_K\! \phi_{t_2}(\vx,\vy) \!\ll_K\! (\vx_p,\vy_q)$. Then, we have $(\vx_r,\vy_s) \!\ll_K\! \phi_{\max\{t_1,t_2\}}(\vx,\vy) \!\ll_K\! (\vx_p,\vy_q)$. Since $\phi_t(\vx_r,\vy_s) \!\rightarrow\! (\hat \vx,\hat \vx)\in S$, and $\phi_t(\vx_p,\vy_q) \rightarrow (\bar \vx,\bar \vx)\in S$, we can once again, due to monotonicity of the system and by invoking a sandwich argument, say that $\phi_{t+\max\{t_1,t_2\}}(\vx,\vy)$ converges to an equilibrium point in $S$ as $t \!\rightarrow\! \infty$. This completes the proof.
\end{IEEEproof}\smallskip

\vspace{-5mm}
\begin{IEEEbiography}[{\includegraphics[width=1in,height=1.25in,clip,keepaspectratio]{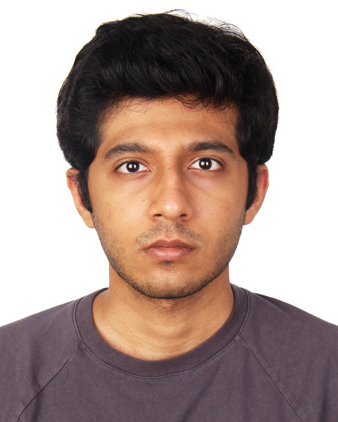}}]%
{Vishwaraj Doshi}
received his B.E. degree in mechanical engineering from the University of Mumbai, Mumbai, MH, India, and Masters degree in Operations Research from North Carolina State University, Raleigh, NC, USA, in 2015 and 2017 respectively. He completed his Ph.D. degree with the Operations Research Graduate Program at North Carolina State University in 2022, and is now a part of the Data Science and Advanced Analytics team at IQVIA. His primary research interests include design of randomized algorithms on graphs, and epidemic models on networks.
\end{IEEEbiography}
\vspace{-5mm}
\begin{IEEEbiography}[{\includegraphics[width=1in,height=1.25in,clip,keepaspectratio]{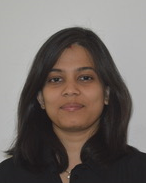}}]%
{Shailaja Mallick}
is a Ph.D. student in the Computer Science Department at North Carolina State University. She received her B.Tech in Computer Science from UCE, Burla, India and Masters in Computer Systems and Networks from Chalmers University of Technology, Sweden. Her current research interests are in the area of social network analysis, network and performance modeling using techniques from mathematical biology, graph theory, stochastic modeling and simulation.
\end{IEEEbiography}
\vspace{-5mm}
\begin{IEEEbiography}[{\includegraphics[width=1in,height=1.25in,clip,keepaspectratio]{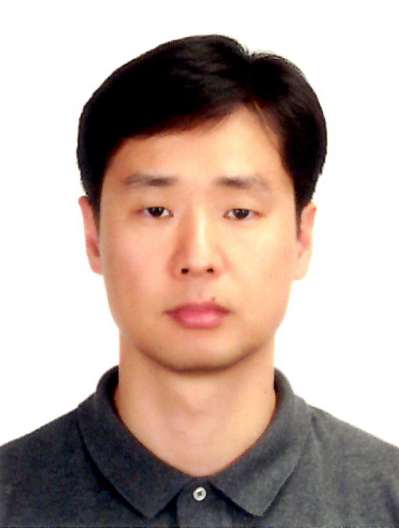}}]%
{Do Young Eun}
(Senior Member, IEEE) received his B.S. and M.S. degree in Electrical Engineering from Korea Advanced Institute of Science and Technology (KAIST), Taejon, Korea, in 1995 and 1997, respectively, and Ph.D. degree from Purdue University, West Lafayette, IN, in 2003. Since August 2003, he has been with the Department of Electrical and Computer Engineering at North Carolina State University, Raleigh, NC, where he is currently a professor. His research interests include distributed optimization for machine learning, machine learning algorithms for networks, distributed and randomized algorithms for large social networks and wireless networks, epidemic modeling and analysis, graph analytics and mining techniques with network applications. He has been a member of Technical Program Committee of various conferences including IEEE INFOCOM, ICC, Globecom, ACM MobiHoc, and ACM Sigmetrics. He is serving on the editorial board of IEEE Transactions on Network Science and Engineering, and previously served for IEEE/ACM Transactions on Networking and Computer Communications Journal, and was TPC co-chair of WASA'11. He received the Best Paper Awards in the IEEE ICCCN 2005, IEEE IPCCC 2006, and IEEE NetSciCom 2015, and the National Science Foundation CAREER Award 2006. He supervised and co-authored a paper that received the Best Student Paper Award in ACM MobiCom 2007.
\end{IEEEbiography}

\end{document}